%% file: TOP-14-018_temp.tex
\pdfoutput=1

\documentclass[11pt,twoside,a4paper,cmspaper,final,collab]{cms-tdr}
\def\svnVersion{333037}\def\cmsCernNoTag{CERN-PH-EP/2013-037}\def\cmsCernDate{\today}\def\cmsMessage{Published in the European Physical Journal C as \href{http://dx.doi.org/10.1140/epjc/s10052-016-3956-5}{\doi{10.1140/epjc/s10052-016-3956-5.}}}

\begin{document}\cmsNoteHeader{TOP-14-018}

\hyphenation{had-ron-i-za-tion}
\hyphenation{cal-or-i-me-ter}
\hyphenation{de-vices}
\RCS$HeadURL: svn+ssh://svn.cern.ch/reps/tdr2/papers/TOP-14-018/trunk/TOP-14-018.tex $
\RCS$Id: TOP-14-018.tex 331909 2016-03-10 15:36:48Z kkousour $
\newlength\cmsFigWidth
\ifthenelse{\boolean{cms@external}}{\setlength\cmsFigWidth{0.85\columnwidth}}{\setlength\cmsFigWidth{0.4\textwidth}}
\ifthenelse{\boolean{cms@external}}{\providecommand{\cmsLeft}{top\xspace}}{\providecommand{\cmsLeft}{left\xspace}}
\ifthenelse{\boolean{cms@external}}{\providecommand{\cmsRight}{bottom\xspace}}{\providecommand{\cmsRight}{right\xspace}}
\cmsNoteHeader{TOP-14-018}

\title{Measurement of the \ttbar production cross section in the all-jets final state in pp collisions at $\sqrt{s}=8$\TeV}

\date{\today}

\abstract{
The cross section for \ttbar production in the all-jets final state is measured in pp collisions at a centre-of-mass energy of 8\TeV at the LHC with the CMS detector, in data corresponding to an integrated luminosity of 18.4\fbinv. The inclusive cross section is found to be $275.6\pm 6.1\stat\pm 37.8\syst\pm 7.2\lum$\unit{pb}. The normalized differential cross sections are measured as a function of the top quark transverse momenta, \pt, and compared to predictions from quantum chromodynamics. The results are reported at detector, parton, and particle levels. In all cases, the measured top quark \pt spectra are significantly softer than theoretical predictions.
}

\hypersetup{%
pdfauthor={CMS Collaboration},%
pdftitle={Measurement of the ttbar production cross section in the all-jets final state in pp collisions at sqrt(s)=8 TeV},%
pdfsubject={CMS},%
pdfkeywords={CMS, physics, top, hadronic}}

\maketitle 

\section{Introduction}

The top quark is an important component of the standard model (SM), especially because of its large mass, and its properties are critical for the overall understanding of the theory. Measurements of the top quark-antiquark pair ($\cPqt\cPaqt$) production cross section test the predictions of quantum chromodynamics (QCD), constrain QCD parameters, and are sensitive to physics beyond the SM. The $\cPqt\cPaqt$ process is also the dominant SM background to many searches for new physical phenomena, and its precise measurement is essential for claiming new discoveries.

The copious top quark data samples produced at the CERN LHC enable measurements of the $\cPqt\cPaqt$ production rate in extended parts of the phase space, and differentially as a function of the kinematic properties of the $\cPqt\cPaqt$ system. Inclusive and differential cross section measurements from proton-proton (pp) collisions at centre-of-mass energies of 7 and 8\TeV have been reported by the ATLAS~\cite{Aad:2010ey,Aad:2011yb,Aad:2012qf,ATLAS:2012aa,Aad:2012mza,Aad:2012vip,Aad:2014kva,Aad:2014zka,Aad:2015eia,Aad:2015pga,Aad:2015dya} and CMS collaborations~\cite{Khachatryan:2010ez,Chatrchyan:2011nb,Chatrchyan:2011ew,Chatrchyan:2011yy,Chatrchyan:2012vs,Chatrchyan:2012bra,Chatrchyan:2012saa,Chatrchyan:2012ria,Chatrchyan:2013kff,Chatrchyan:2013ual,Chatrchyan:2013faa,Khachatryan:2014loa,Khachatryan:2015oqa}. These are significantly more precise than the measurements of $\cPqt\cPaqt$ production in proton-antiproton collisions performed at the Tevatron~\cite{Aaltonen:2013wca}. In this paper, we report new results from pp collision data at $\sqrt{s}=8\TeV$, collected with the CMS detector. Measurements of the $\cPqt\cPaqt$ inclusive cross section and the normalized differential cross sections are presented for the first time in the all-jets final state at this collision energy. The results are compared to QCD predictions, and are in agreement with other measurements in different decay channels.

Top quarks decay almost exclusively into a W boson and a b quark. Events in which both W bosons from
the $\cPqt\cPaqt$ decay produce a pair of light quarks constitute the so-called  all-jets channel. As a result, the final state consists of at least six partons (more are possible from initial- and final-state radiation), two of which are b quarks. Despite the large number of combinatorial possibilities, it is possible to fully reconstruct the kinematical properties of the $\cPqt\cPaqt$ decay products, unlike in the leptonic channels where the presence of one or two neutrinos makes the full event interpretation ambiguous. However, the presence of a large background from multijet production, and the larger number of jets in the final state make the measurement of the $\cPqt\cPaqt$ cross section in the all-jets final state more uncertain compared to the leptonic channels. 
Nevertheless, a high-purity signal sample can be selected, which increases significantly the signal-over-background ratio compared to previous measurements in this decay channel~\cite{Chatrchyan:2013ual,Aaltonen:2010pe,Abazov:2009ss}.

\section{The CMS detector}\label{sec:det}

The central feature of the CMS apparatus is a superconducting solenoid of 6\unit{m} internal diameter, providing a magnetic field of 3.8\unit{T}. Within the solenoid volume are a silicon pixel and strip tracker, a lead tungstate crystal electromagnetic calorimeter, and a brass and scintillator hadron calorimeter. Extensive forward calorimetry (pseudorapidity $\abs{\eta}>3.0$) complements the coverage provided by the barrel ($\abs{\eta}<1.3$) and endcap ($1.3<\abs{\eta}<3.0$) detectors. Muons are measured in gas-ionization detectors embedded in the steel flux-return yoke outside the solenoid. The first level of the CMS trigger system, composed of custom hardware processors, uses information from the calorimeters and muon detectors to select the most interesting events in a fixed time interval of less than 4\mus. The high-level trigger (HLT) processor farm further decreases the event rate from around 100\unit{kHz} to around 300\unit{Hz}, before data storage. A detailed description of the CMS apparatus, together with the definition of the coordinate system used and the relevant kinematic variables, can be found in Ref.~\cite{Chatrchyan:2008zzk}.

\section{Event simulation}\label{sec:sim}

{\tolerance=600
The $\cPqt\cPaqt$ events are simulated using the leading-order (LO) \MADGRAPH (v.~5.1.5.11) event generator~\cite{Alwall:2014hca}, which incorporates spin correlations through the \textsc{madspin}~\cite{Artoisenet:2012st} package and the simulation of up to three additional partons. The value of the top quark mass is set to $m_{\cPqt}=172.5\GeV$ and the proton structure is described by the parton distribution functions (PDFs) from CTEQ6L1~\cite{cteq}. The generated events are subsequently processed with \PYTHIA (v.~6.426)~\cite{Sjostrand:2006za} which utilizes tune Z2* for parton showering and hadronization, and the MLM prescription~\cite{MLM} is used for matching of matrix element jets to those from parton shower. The \PYTHIA Z2* tune is derived from the Z1* tune~\cite{Field:2010bc}, which uses the CTEQ5L PDF~\cite{cteq}, whereas Z2* adopts CTEQ6L~\cite{cteq}. The CMS detector response is simulated using \GEANTfour (v.~9.4)~\cite{geant}.
\par}

In addition to the \MADGRAPH simulation, predictions obtained with the next-to-leading-order (NLO) generators \MCATNLO (v.~3.41)~\cite{mcatnlo} and \POWHEG (v.~1.0 r1380)~\cite{Nason:2004rx} are also compared to the measurements. While \POWHEG and \MCATNLO are formally equivalent up to NLO accuracy, they differ in the techniques used to avoid double counting of the radiative corrections when interfacing with the parton shower generators. Two different \POWHEG samples are used: one uses \PYTHIA and the other \HERWIG (v.~6.520)~\cite{HERWIG} for parton showering and hadronization. The events generated with \MCATNLO are interfaced with \HERWIG. The \HERWIG AUET2 tune~\cite{auet2tune} is used to model the underlying event in the \POWHEG{}+\HERWIG sample, while the default tune is used in the \MCATNLO{}+\HERWIG sample. The proton structure is described by the PDF sets CT10~\cite{CT10} and CTEQ6M~\cite{cteq} for \POWHEG and \MCATNLO, respectively. The QCD multijet events are simulated using \MADGRAPH (v. 5.1.3.2) interfaced with \PYTHIA (v.~6.424).

\section{Event reconstruction and selection}

\subsection{Jet reconstruction}

Jets are reconstructed with the anti-\kt clustering algorithm~\cite{Cacciari:2005hq,Cacciari:2008gp} with a distance parameter of 0.5. The input to the jet clustering algorithm is the collection of particle candidates that are reconstructed with the particle-flow (PF) algorithm~\cite{CMS-PAS-PFT-09-001,CMS-PAS-PFT-10-001}. In the PF event reconstruction all stable particles in the event, \ie electrons, muons, photons, and charged and neutral hadrons, are reconstructed as PF candidates using a combination of all of the subdetector information to obtain an optimal determination of their directions, energies, and types. All the reconstructed vertices in the event are ordered according to the sum of squared transverse momenta (\pt) of tracks used to reconstruct it and the vertex with the largest sum is considered the primary one, while all the rest are considered as pileup vertices. In order to mitigate the effect of multiple interactions in the same bunch crossing (pileup), charged PF candidates that are unambiguously associated with pileup vertices are removed prior to the jet clustering. This procedure is called charged-hadron subtraction (CHS)~\cite{JES}. An offset correction is applied for the additional energy inside of the jet due to neutral hadrons or photons from pileup. The resulting jets require a small residual energy correction, mostly due to the thresholds for reconstructed tracks and clusters in the PF algorithm and reconstruction inefficiencies~\cite{JES}.

The identification of jets that likely originate from the hadronization of b quarks is done with the ``combined secondary vertex" (CSV) b tagger~\cite{Chatrchyan:2012jua}. The CSV algorithm combines the information from track impact parameters and identified secondary vertices within a given jet, and provides a continuous discriminator output.

\subsection{Trigger}\label{sec:trig}

The data used for this measurement were collected with a multijet trigger event selection (path) which, from the HLT, required at least four jets reconstructed from calorimetric information with a \pt threshold of $50\GeV$ and $\abs{\eta}<3.0$. The hardware trigger required the presence of two central ($\abs{\eta}<3.0$) jets above various \pt thresholds (52--64\GeV), or the presence of four central jets with lower \pt thresholds (32--40\GeV), or the scalar sum of all jets \pt to be greater than 125 or 175\GeV. The various thresholds were adjusted within the quoted ranges according to the instantaneous luminosity. The trigger paths employed were unprescaled for a larger part of the run, yielding a data sample corresponding to an integrated luminosity of 18.4\fbinv.

\subsection{Selection and kinematic top quark pair reconstruction}\label{sec:Sel}

Selected events are required to contain at least six reconstructed jets with $\pt>40\GeV$ and $\abs{\eta}<2.4$ (jets are required to be within the tracker acceptance in order to apply the CHS), with at least four of the jets having $\pt>60\GeV$ (so that the trigger efficiency is greater than 80\% and the data-to-simulation correction factor smaller than 10\%). Among the six jets with the highest \pt (leading jets), at least two must be identified as coming from b hadronization by the CSV algorithm at the medium working point (CSVM), with a typical b quark identification efficiency of 70\% and misidentification probability for light quarks of 1.4\%, and these are considered the most probable b jet candidates. If there are more than two such jets, which happens in approximately 2\% of the events, then the two with the highest \pt are chosen. To select events compatible with the $\cPqt\cPaqt$ hypothesis, and to improve the resolution of the reconstructed quantities, a kinematic fit is performed that utilizes the constraints of the $\cPqt\cPaqt$ decay. A $\chi^2$ fit is performed, starting with the reconstructed jet four-momenta, which are varied within their experimental \pt and angular resolutions, imposing a W boson mass constraint (80.4\GeV~\cite{Agashe:2014kda}) on the light-quark pairs, and requiring that the top quark and antiquark have equal mass. Out of all the possible combinations from the six input jets, the algorithm returns the one with the smallest $\chi^2$ and the resulting parton four momenta, which are used to compute the reconstructed top quark mass ($m_{\cPqt}^\text{rec}$). The probability of the converged kinematic fit is required to be greater than 0.15. Overall, the kinematic fit requirements select approximately 5\% (2\%) of the $\cPqt\cPaqt$ (background) events. The distance in the $\eta$--$\phi$ space between the two b quark candidates must be $\Delta R_{\cPqb\cPqb}=\sqrt{\smash[b]{\left(\Delta\eta_{\cPqb\cPqb}\right)^2+\left(\Delta\phi_{\cPqb\cPqb}\right)^2}}>2.0$, which has an efficiency of roughly 75\% (50\%) on $\cPqt\cPaqt$ (background) events. The last two requirements are applied to select events with unambiguous top quark pair interpretation and to suppress the QCD background that originates from gluon splitting into collinear b quarks~\cite{Chatrchyan:2013xza}.

\section{Signal extraction}\label{sec:Signal}

\begin{figure}[hbt]
  \centering
    \includegraphics[width=0.49\textwidth]{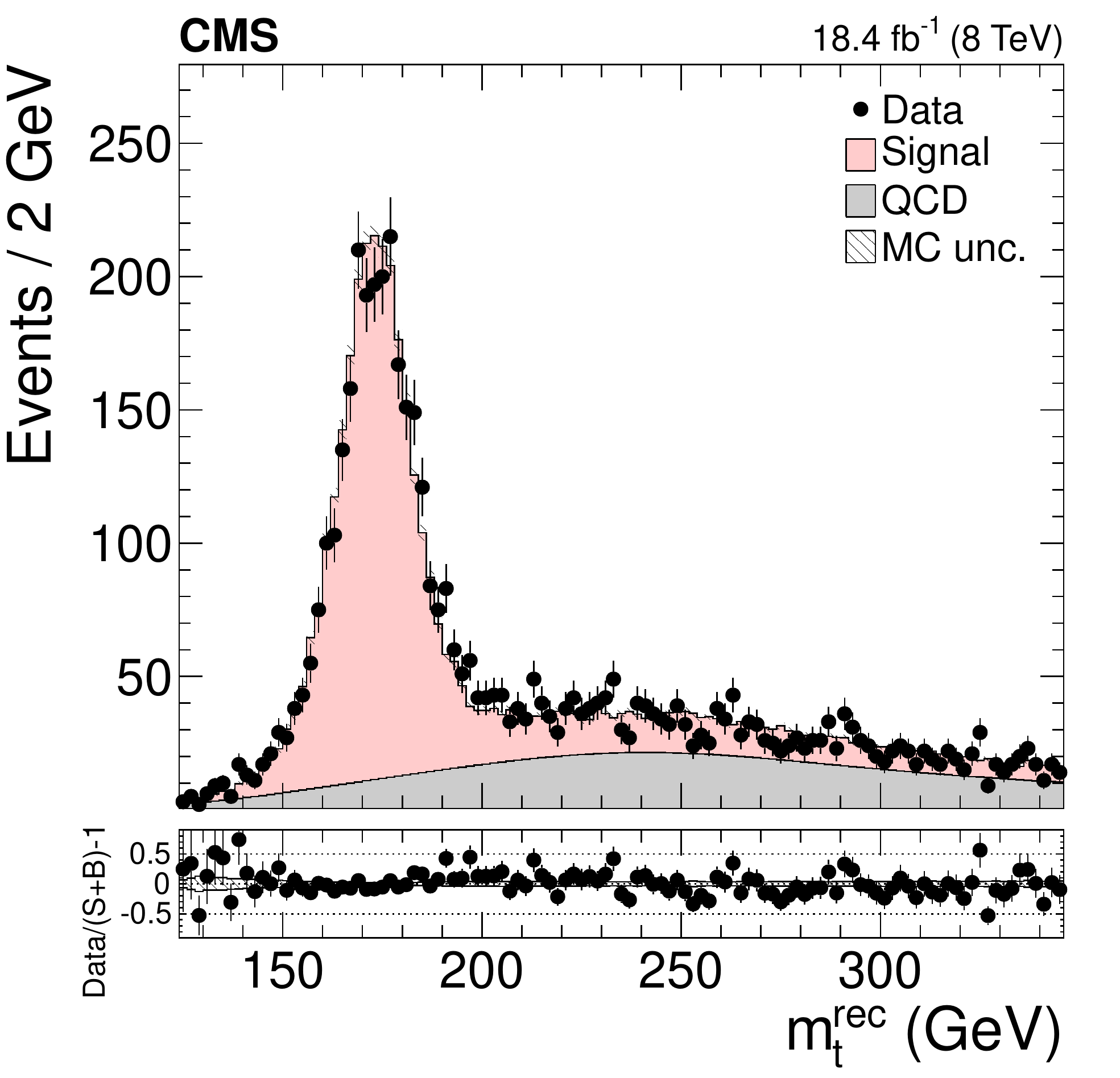}
    \caption{Distribution of the reconstructed top quark mass after the kinematic fit. The normalizations of the $\cPqt\cPaqt$ signal and the QCD multijet background are taken from the template fit to the data. The bottom panel shows the fractional difference between the data and the sum of signal and background predictions, with the shaded band representing the MC statistical uncertainty.}
    \label{fig:mtop}
  \end{figure}

\begin{figure}[hbt]
  \centering
    \includegraphics[width=0.49\textwidth]{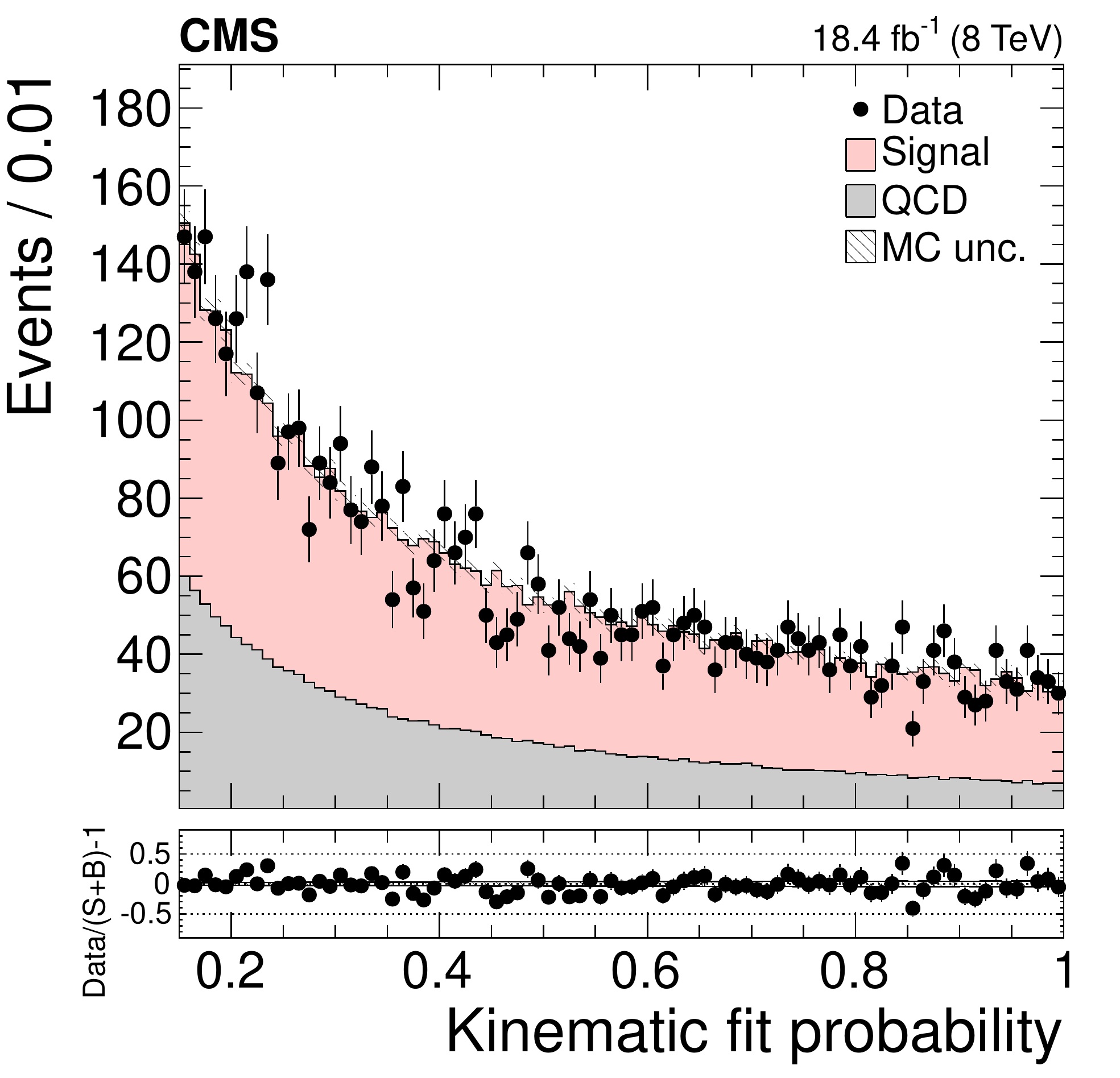}
    \includegraphics[width=0.49\textwidth]{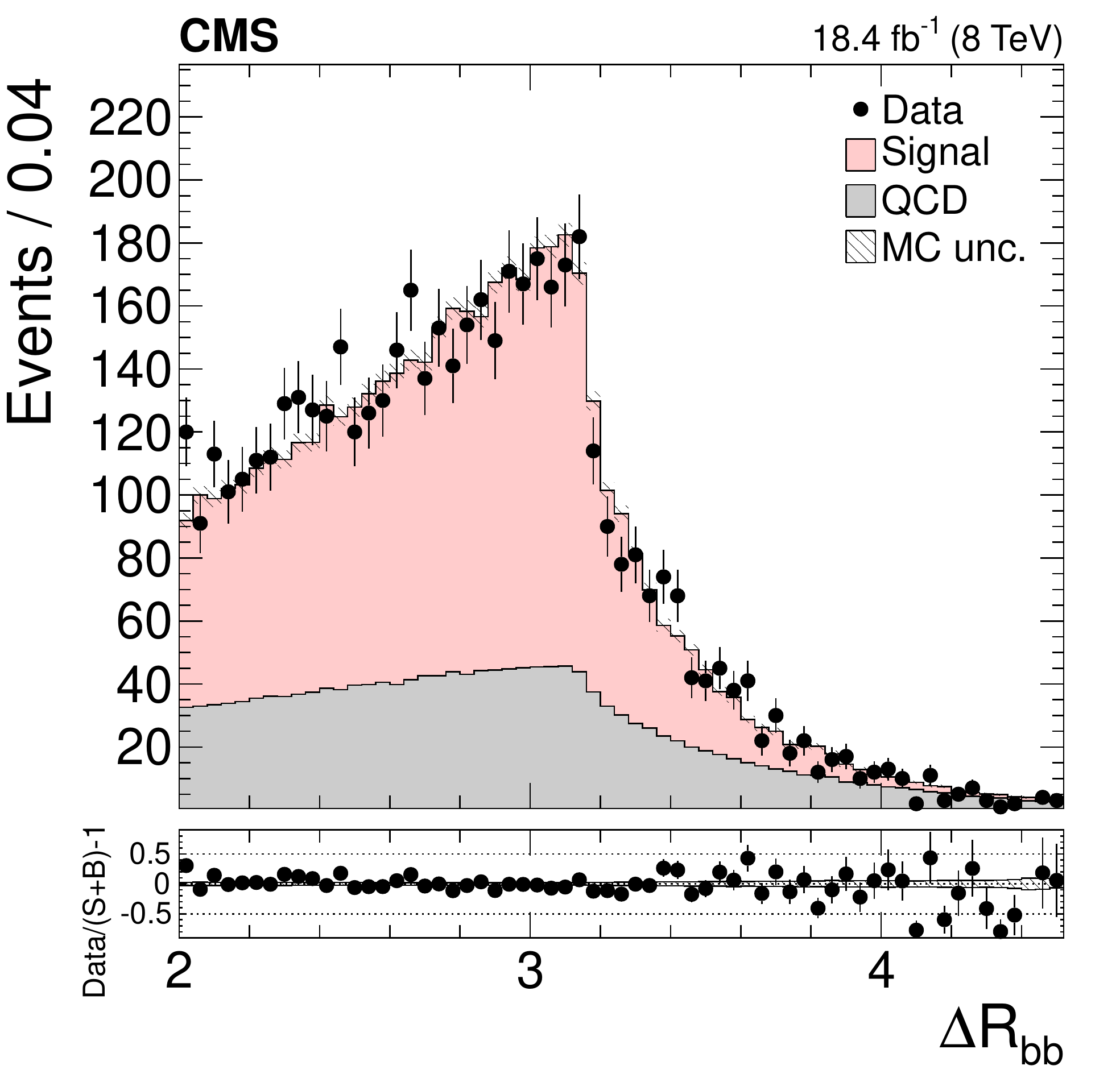}
    \caption{Distribution of the kinematic fit probability (\cmsLeft). Distribution of the distance between the reconstructed b partons in the $\eta$--$\phi$ plane (\cmsRight). The normalizations of the $\cPqt\cPaqt$ signal and the QCD multijet background are taken from the template fit to the data. The bottom panels show the fractional difference between the data and the sum of signal and background predictions, with the shaded band representing the MC statistical uncertainty.}
    \label{fig:properties}
  \end{figure}

\begin{figure*}[hbtp]
  \centering
    \includegraphics[width=0.45\textwidth]{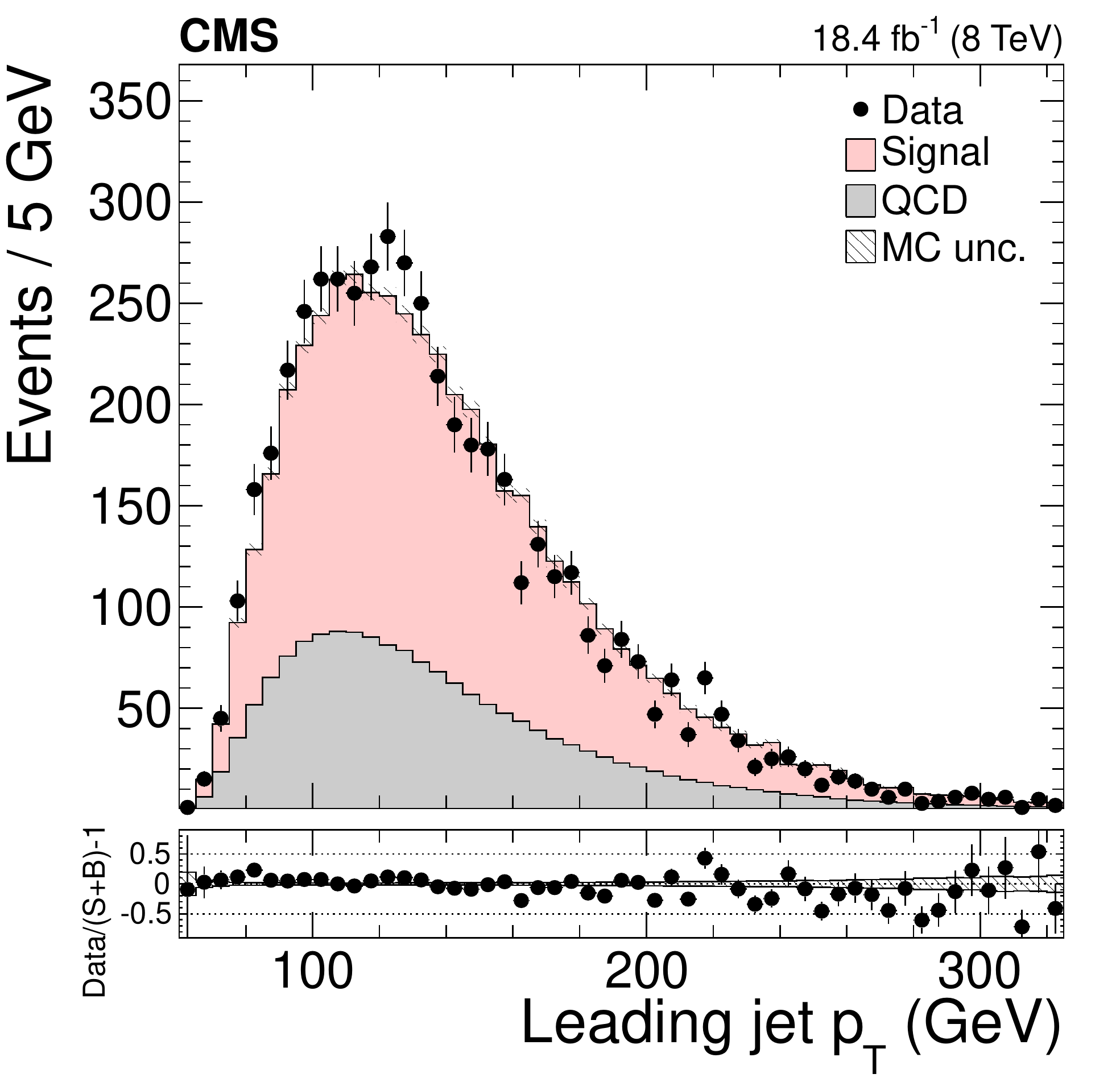}
    \includegraphics[width=0.45\textwidth]{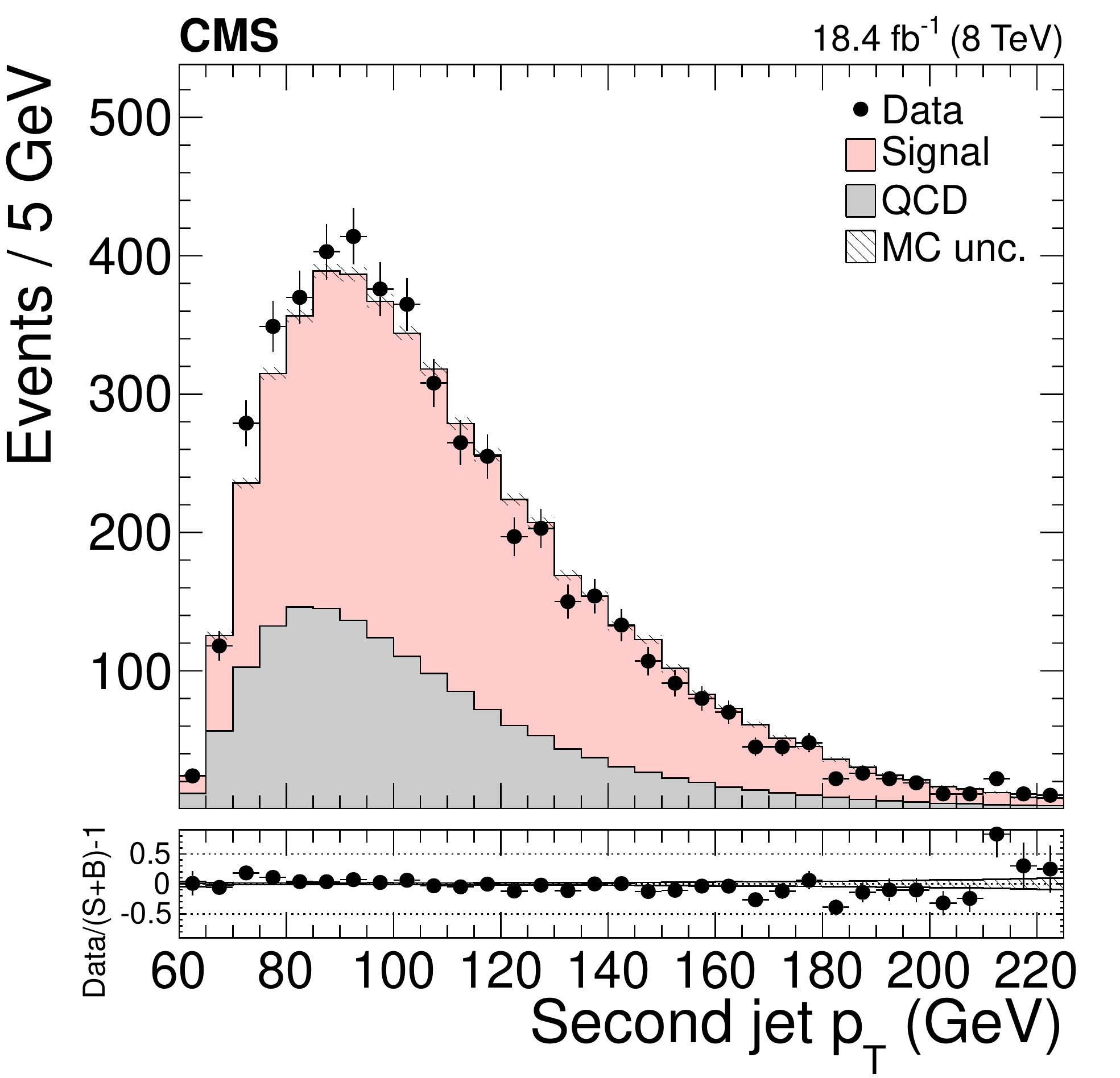}
    \includegraphics[width=0.45\textwidth]{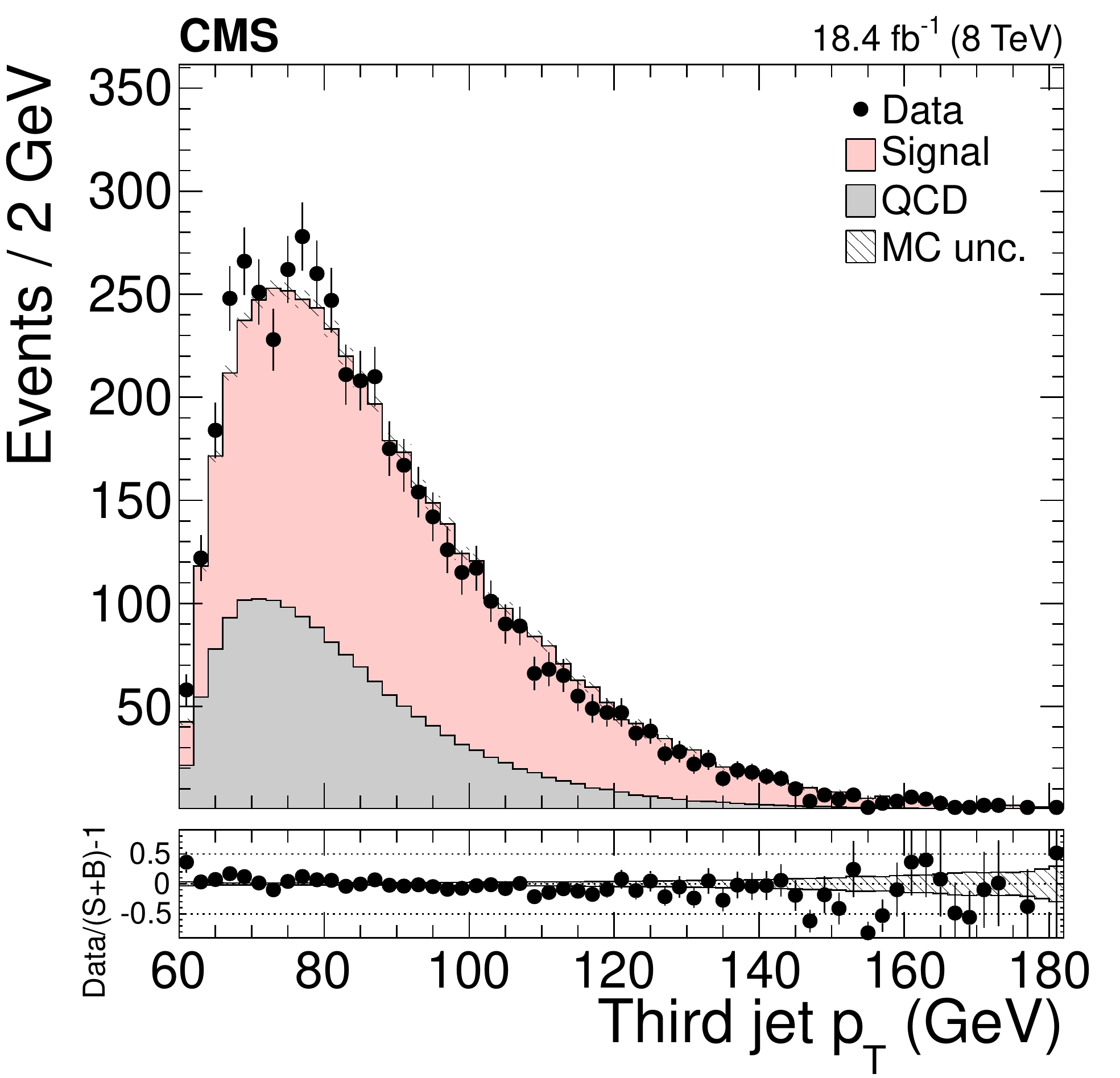}
    \includegraphics[width=0.45\textwidth]{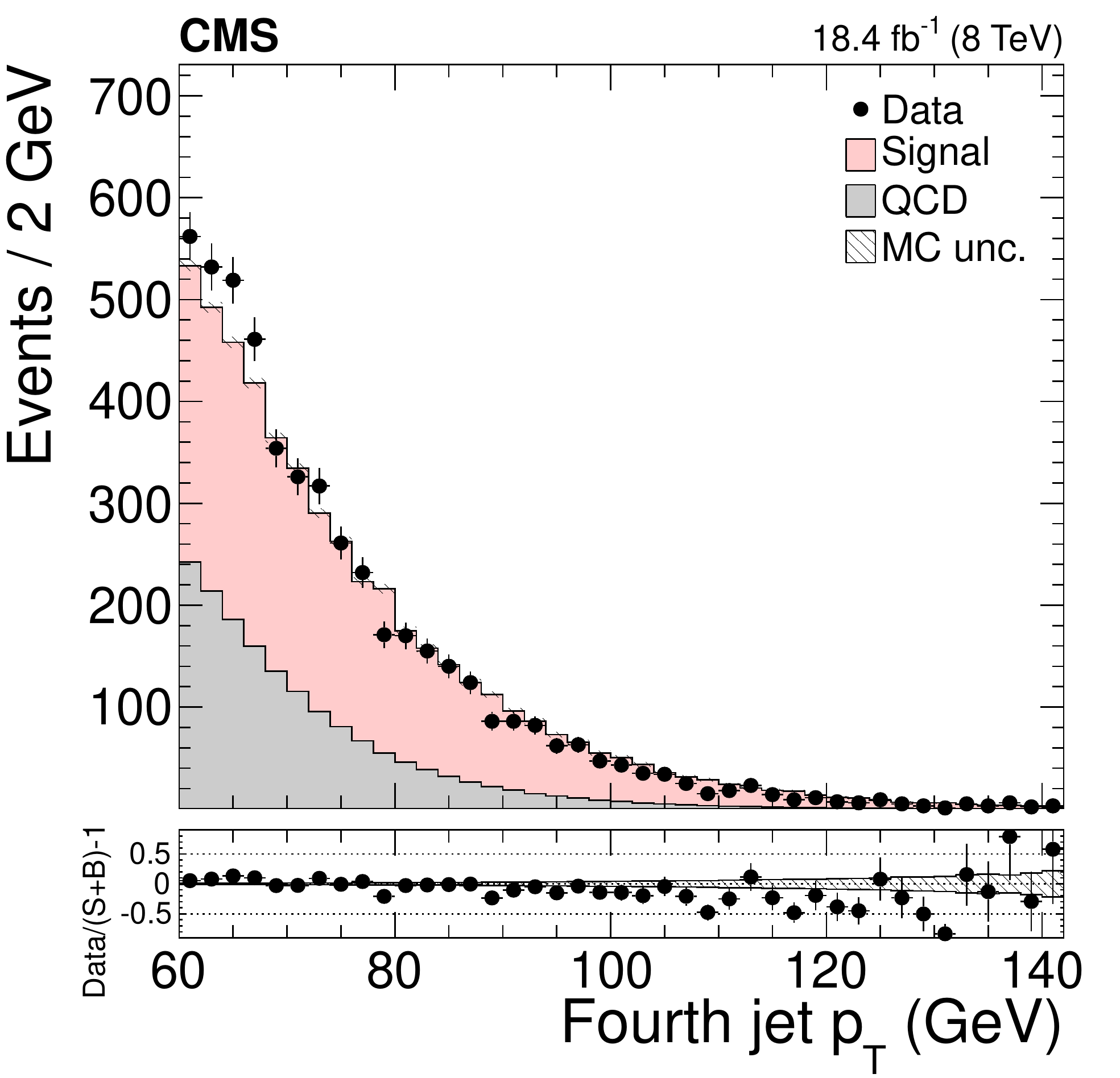}
    \includegraphics[width=0.45\textwidth]{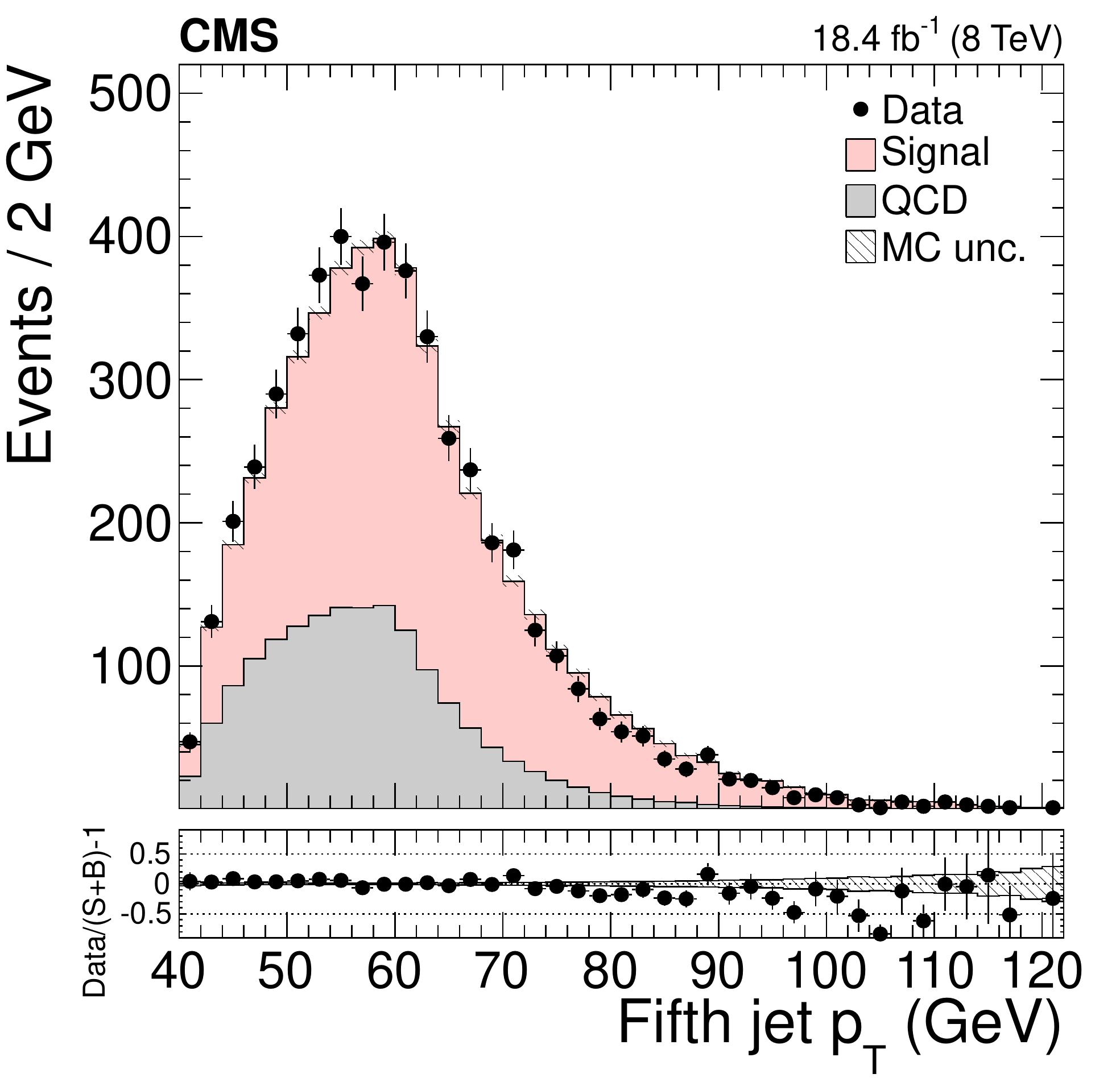}
    \includegraphics[width=0.45\textwidth]{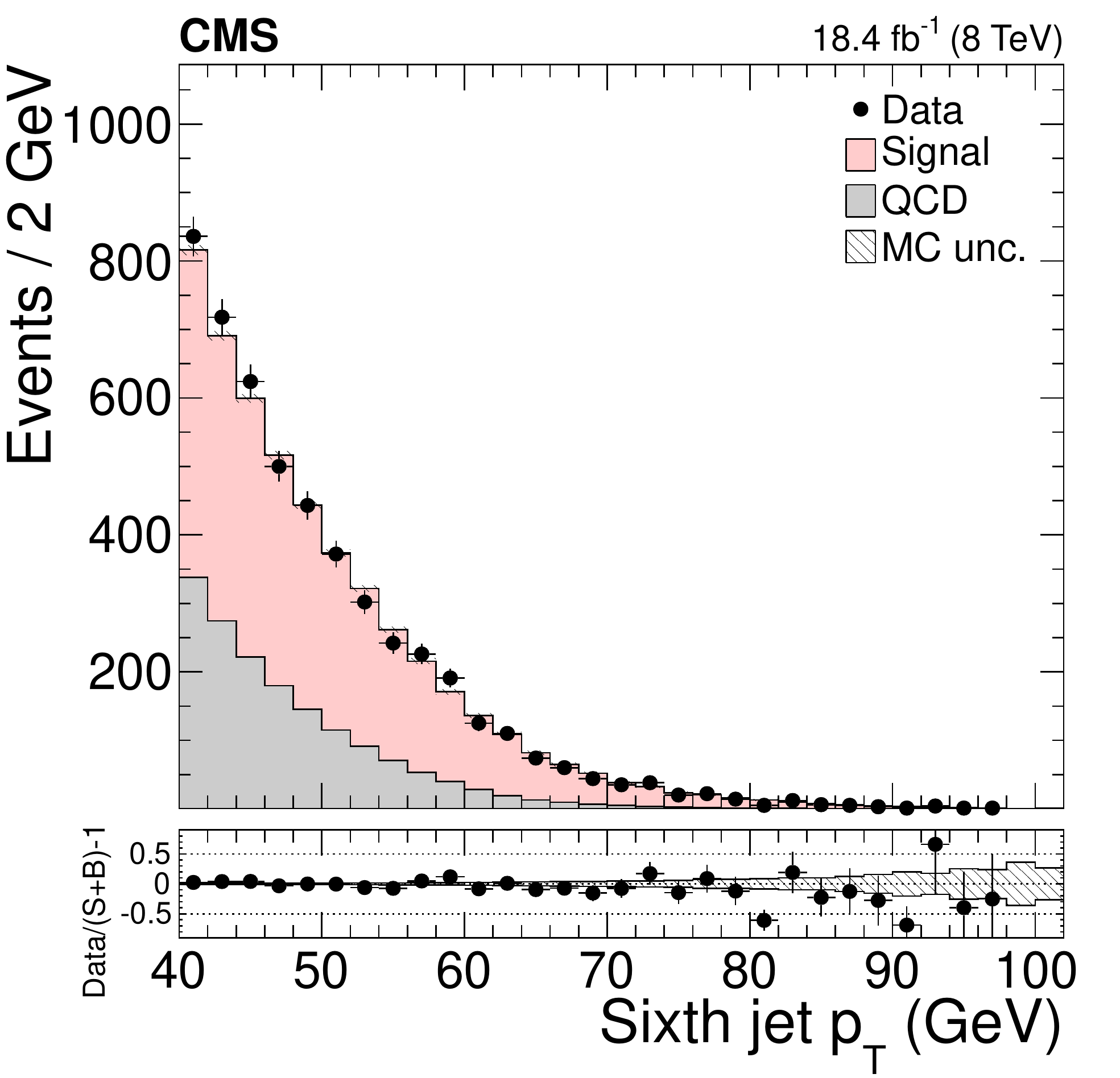}
    \caption{Distribution of the \pt of the six leading jets. The normalizations of the $\cPqt\cPaqt$ signal and the QCD multijet background are taken from the template fit to the data. The bottom panels show the fractional difference between the data and the sum of signal and background predictions, with the shaded band representing the MC statistical uncertainty.}
    \label{fig:jetPt}
  \end{figure*}

\begin{figure}[hbt]
  \centering
    \includegraphics[width=0.49\textwidth]{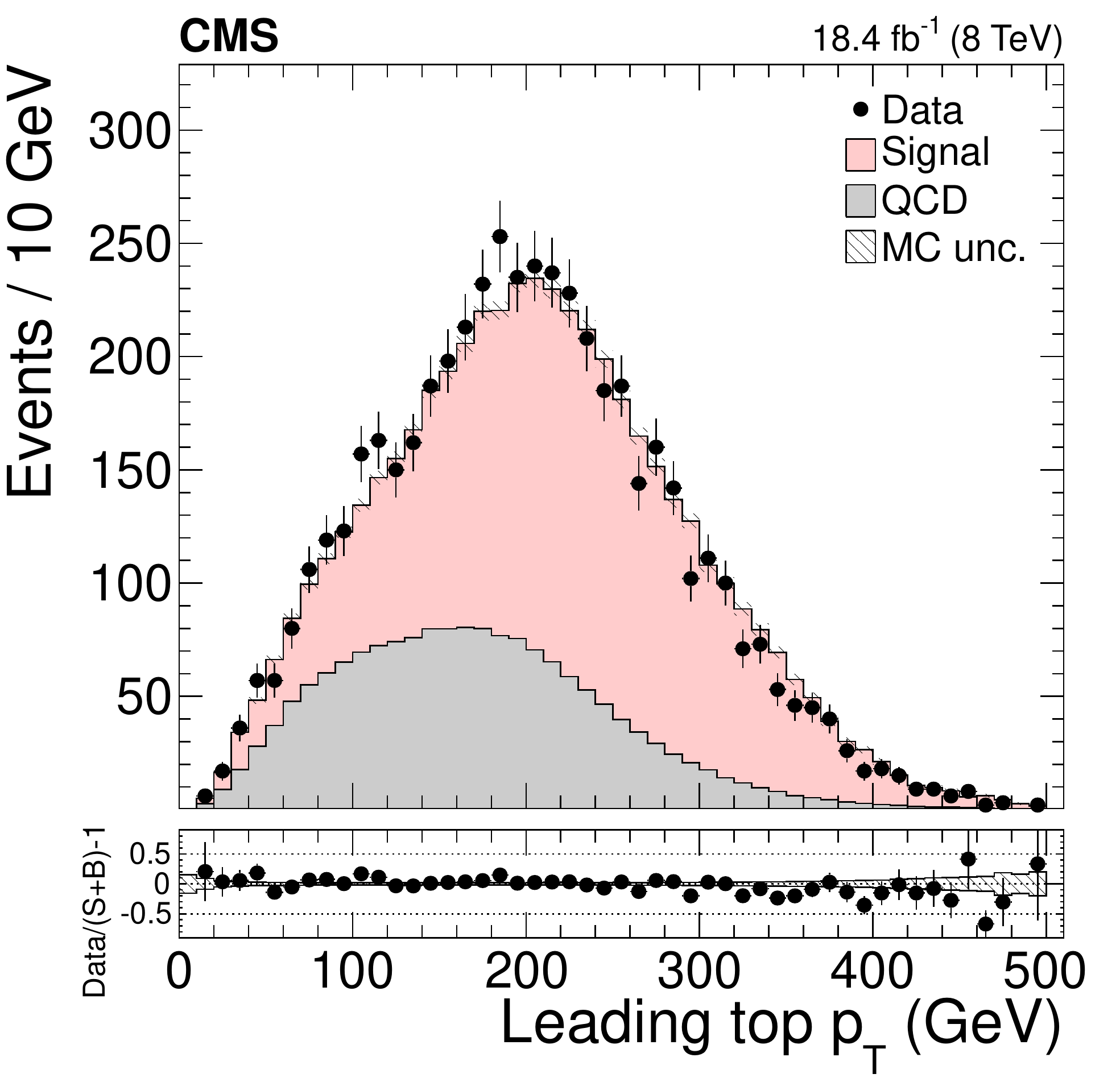}
    \includegraphics[width=0.49\textwidth]{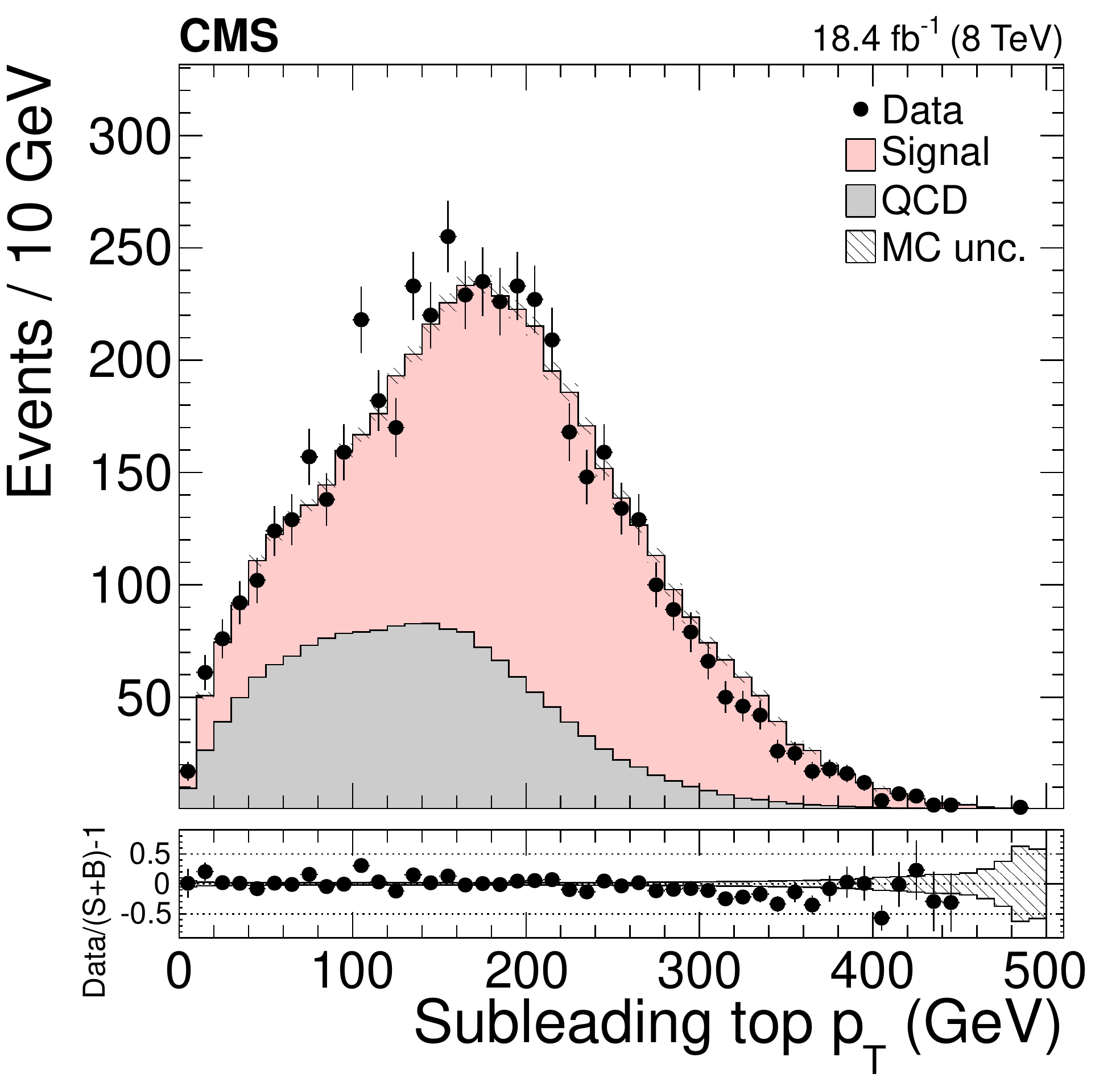}
    \caption{Distribution of the leading (\cmsLeft) and subleading (\cmsRight) reconstructed top quark \pt. The normalizations of the $\cPqt\cPaqt$ signal and the QCD multijet background are taken from the template fit to the data. The bottom panels show the fractional difference between the data and the sum of signal and background predictions, with the shaded band representing the MC statistical uncertainty.}
    \label{fig:ptTop}
  \end{figure}

\begin{figure}[hbt]
  \centering
    \includegraphics[width=0.49\textwidth]{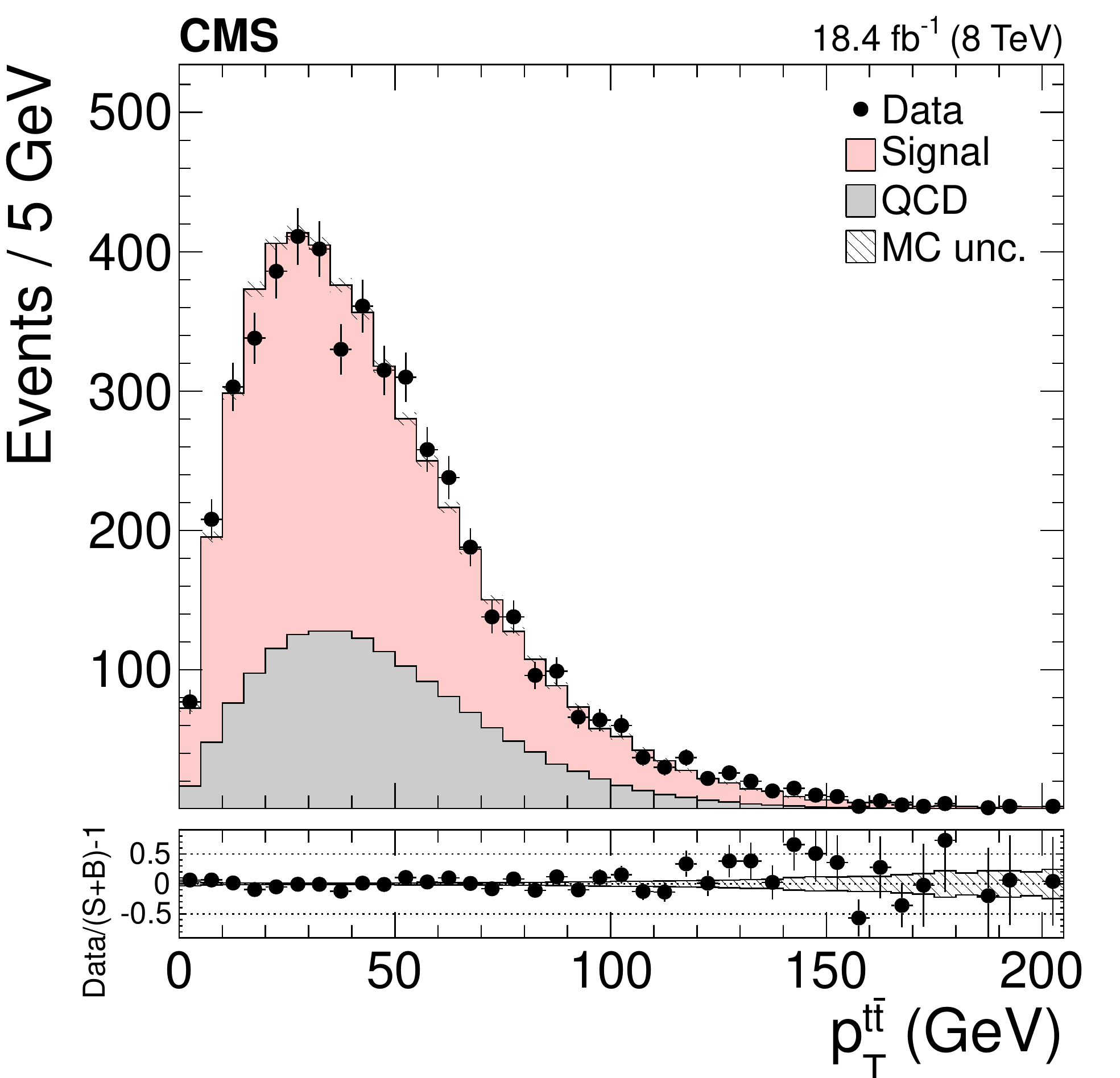}
    \includegraphics[width=0.49\textwidth]{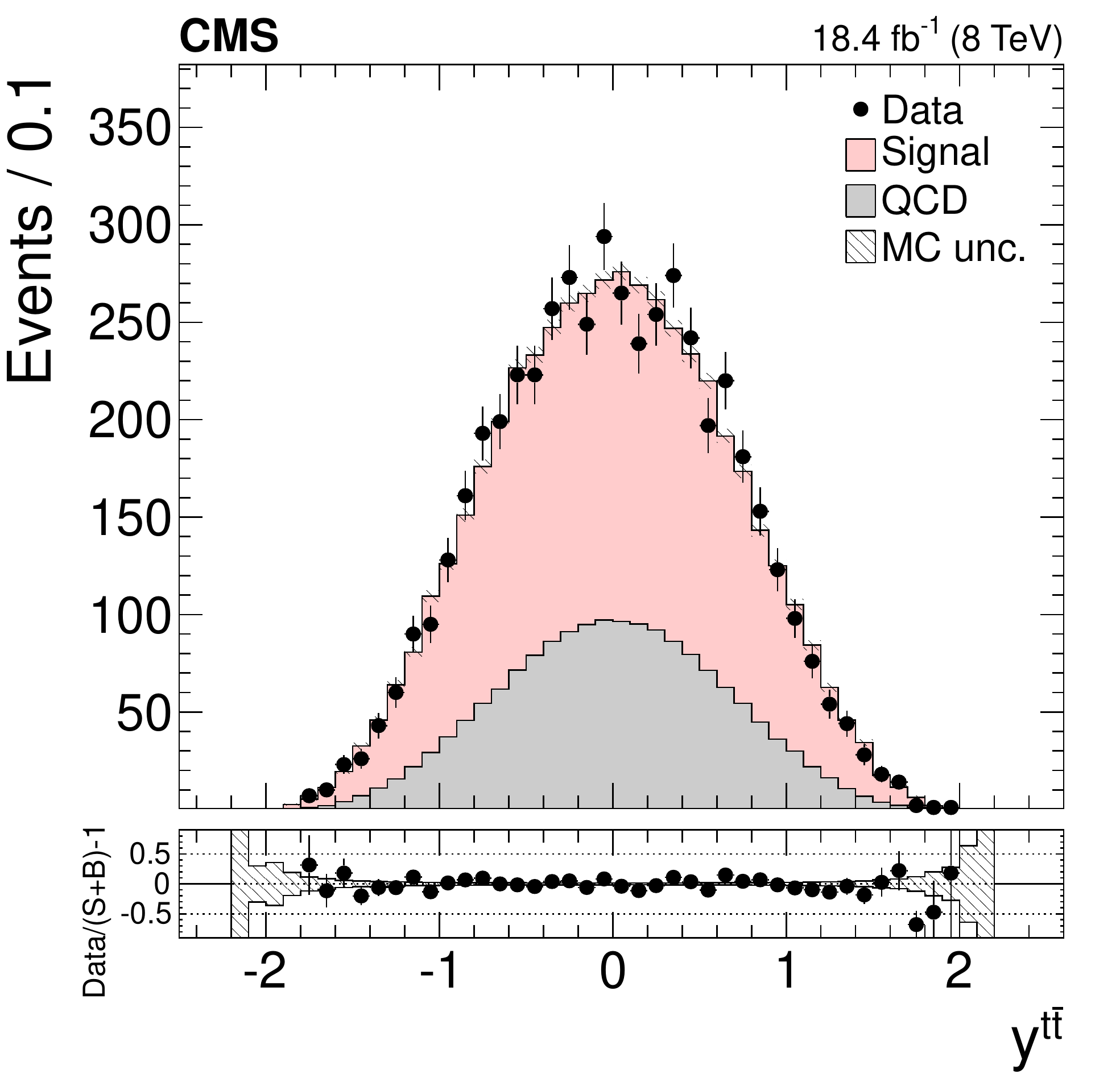}
    \caption{Distribution of the \pt (\cmsLeft) and the rapidity (\cmsRight) of the reconstructed top quark pair. The normalizations of the $\cPqt\cPaqt$ signal and the QCD multijet background are taken from the template fit to the data. The bottom panels show the fractional difference between the data and the sum of signal and background predictions, with the shaded band representing the MC statistical uncertainty.}
    \label{fig:ttbar}
  \end{figure}

The background to the $\cPqt\cPaqt$ signal is dominated by the QCD multijet production process, while the other backgrounds, such as the associated production of vector bosons with jets, are negligible. Due to the limited size of the Monte Carlo (MC) simulated samples, the background is determined directly from the data. A QCD-dominated event sample is selected with the trigger and offline requirements described in Section~\ref{sec:Sel} and requiring zero CSVM b tagged jets. In these events the most probable b quark candidates are determined by the kinematic fit. The resulting sample contains a negligible fraction of $\cPqt\cPaqt$ events ($<1$\%) and is treated exactly like the signal sample. After applying the $\Delta R_{\cPqb\cPqb}>2.0$ and the fit probability requirements, the reconstructed top-like kinematic properties of events with no b jet are very similar to those with two b jets (confirmed using simulated QCD events). We use this QCD-dominated control sample to extract the shape (templates) of the various kinematic observables. The number of $\cPqt\cPaqt$ events (signal yield) is extracted from a template fit of $m_{\cPqt}^\text{rec}$ to the data using parametrized shapes for signal and background distributions, where the signal shape is taken from the $\cPqt\cPaqt$ simulation and the QCD shape is taken from the control data sample described above. The background and signal yields are determined via a maximum likelihood fit to the $m_{\cPqt}^\text{rec}$ distribution and are used to normalize the corresponding samples. Figures~\ref{fig:mtop} and~\ref{fig:properties} show the fitted mass and the kinematic fit probability and $\Delta R_{\cPqb\cPqb}$ distributions. The \pt distribution of the six leading jets is shown in Fig.~\ref{fig:jetPt}. From the output of the kinematic fit one can reconstruct the two top quark candidates, whose \pt are shown in Fig.~\ref{fig:ptTop}, and the properties of the $\cPqt\cPaqt$ system (\pt, rapidity $y$) are shown in Fig.~\ref{fig:ttbar}. Overall, the data sample is dominated by signal events, and the data are in agreement with the fit results. The jet \pt spectra in data appear to be systematically softer than in the simulation, in agreement with the observations in Ref.~\cite{Khachatryan:2015oqa}, related to a softer measured top quark \pt spectrum.

\section{Systematic uncertainties}\label{sec:unc}

The measurement of the $\cPqt\cPaqt$ cross section is affected by several sources of systematic uncertainty, both experimental and theoretical, which are described below and summarized in Table~\ref{tab:systematics}. The quoted values refer to the inclusive measurement, with small variations observed in the bins of the differential measurement presented in Section~\ref{sec:diffxsec}.

\begin{itemize}
\item \textbf{Background modeling:} the QCD $m_{\cPqt}^\text{rec}$ template shape derived from the data control sample is varied according to the uncertainty of the method evaluated with simulated events, which impacts the extracted signal yield moderately (4.9\%).
\item \textbf{Trigger efficiency:} the efficiency of the trigger path is taken from the simulation and corrected with an event-by-event scale factor ($\mathrm{SF}_\text{trig}$), calculated from data independent samples, that depends on the fourth jet \pt. In the phase space of the measurement, the $\mathrm{SF}_\text{trig}$ is greater than 0.83 and on average 0.96. The associated uncertainty is conservatively defined as $(1-\mathrm{SF}_\text{trig})/2$ and has a small impact (2.0\%) on the cross section.
\item \textbf{Jet energy scale and resolution:} the jet energy scale (JES) and jet resolution (JER) uncertainties have significant impacts on the measured cross section due to the relatively high \pt requirements on the fourth and sixth of the leading jets. In the simulated events, jets are shifted (smeared) according to the \pt- and $\eta$-dependent JES (JER) uncertainty, prior to the kinematic fit, and the full event interpretation is repeated. The JES (JER) has a dominant (small) effect on the cross section measurement of 7.0\% (3.5\%). In addition, the JES/JER uncertainties affect the signal template, with a negligible impact ($\approx$1\%) on the cross section measurement.
\item \textbf{ b tagging:} the performance of the b tagger has a dominant effect on the signal acceptance because the selected events are required to have at least two jets satisfying the CSVM requirement. An event-by-event scale factor ($\mathrm{SF}_\text{btag}$) is applied to the simulation, which accounts for the discrepancies between data and simulation in the efficiency of tagging true b jets and in the misidentification rate~\cite{Chatrchyan:2012jua}. The average value of $\mathrm{SF}_\text{btag}$ is 0.99. The uncertainty in the $\mathrm{SF}_\text{btag}$ is taken into account by weighting each event with the shifted value of $\mathrm{SF}_\text{btag}$ which results in a cross section uncertainty of 7.3\%. This is the leading systematic uncertainty.
\item \textbf{Integrated luminosity:} the uncertainty on the integrated luminosity is estimated to be 2.6\%~\cite{CMS-PAS-LUM-13-001}.
\item \textbf{Matching partons to showers:} the impact of the choice of the scale that separates the description of jet production via matrix elements or parton shower in \MADGRAPH is studied by changing its reference value of 20 to 40 and 10\GeV, resulting in an asymmetric effect of $-4.2,\,+2.4\%$ on the cross section.
\item \textbf{Renormalization and factorization scales: } the uncertainty in modelling of the hard-production process is assessed through changes in the renormalization and factorization scales in the \MADGRAPH sample by factors of two and half, relative to their common nominal value, which is set to the $Q$ of the hard process. In \MADGRAPH, $Q$ is defined by $Q^2 = m^2_{\cPqt} + \Sigma p^2_{\mathrm{T}}$, where the sum is over all additional final state partons in the matrix element calculations. The effect on the measured cross section is moderate and asymmetric ($-0.5,\,+3.8\%$).
\item \textbf{Parton distribution functions: } following the PDF4LHC prescription~\cite{Alekhin:2011sk,Botje:2011sn}, the uncertainty on the cross section is estimated to be 1.5\%, taking the largest deviation on the signal acceptance from all the considered PDF eigenvectors. 
\item \textbf{Non-perturbative QCD: } the impact of non-perturbative QCD effects is estimated by studying various tunes of the \PYTHIA shower model that predict different underlying event (UE) activity and strength of the color reconnection (CR), namely, the Perugia 2011, Perugia 2011 mpiHi, and Perugia 2011 Tevatron tunes, described in Ref.~\cite{Skands:2010ak}, were used. The effect on the measured cross section is moderate: 4.4\% for the UE and 1.4\% for the CR.
\item \textbf{Hadronization model: } the effect of the hadronization model on the signal efficiency is estimated by comparing the predictions from the \MCATNLO+\HERWIG and \POWHEG+\PYTHIA simulations, and it amounts to 2\%.
\end{itemize}

\begin{table}[htb]
\centering
\topcaption{Fractional uncertainties in the inclusive $\cPqt\cPaqt$ production cross section. \label{tab:systematics}}
\begin{tabular}{lr}
 \hline
 Source				&			\\
 \hline
 Background modeling   		& 	$\pm 4.9\%$	\\
 JES            		& 	$-7.0,\,+6.8\%$	\\
 JER            		& 	$\pm 3.5\%$	\\
 b tagging         		& 	$\pm 7.3\%$	\\
 Trigger efficiency 		& 	$-2.2,\,+2.0\%$	\\
 Underlying event		& 	$\pm 4.4\%$	\\
 Matching partons to showers	& 	$-4.2,\,+2.4\%$	\\
 Factorization and renormalization scales	& 	$-0.5,\,+3.8\%$	\\
 Color reconnection		&	$\pm 1.4\%$	\\
 Parton distribution function   &	$\pm 1.5\%$	\\
 Hadronization                  &       $\pm 2.0\%$	\\
 \hline
 Total systematic uncertainty	&	${\pm}13.7\%$	\\
 Statistical uncertainty        &	$\pm 2.3\%$	\\
 Integrated luminosity		&	$\pm 2.6\%$	\\
 \hline
\end{tabular}
\end{table}

\section{Results}

\subsection{Inclusive cross section}

The signal yield ($N_{\cPqt\cPaqt}$), extracted as described in Section~\ref{sec:Signal}, is used to compute the inclusive $\cPqt\cPaqt$ production cross section, according to the formula
\begin{equation}\label{eq:xsec}
\sigma_{\cPqt\cPaqt}=\frac{N_{\cPqt\cPaqt}}{\left(\mathcal{A}\epsilon\right)\,\mathcal{L}},
\end{equation}

where $(\mathcal{A}\,\epsilon)$ is the simulated signal acceptance times efficiency in the measurement phase space (${\approx}7\times 10^{-4}$) corrected event-by-event with the trigger and b tagging efficiency scale factors and $\mathcal{L}$ is the integrated luminosity. The fitted signal amounts to $3416\pm 79$ events. Taking into account the systematic uncertainties discussed in Section~\ref{sec:unc}, the measured cross section is
\begin{equation}
\sigma_{\cPqt\cPaqt}=275.6\pm 6.1\stat\pm 37.8\syst\pm 7.2\lum\unit{pb}.
\end{equation}

The precision of the measured inclusive cross section is dominated by the systematic uncertainties, and in particular by those related to JES and b tagging.

In order to parametrize the dependence of the result on the top quark mass assumption, the measurement was repeated using signal simulated samples with different generated top quark masses (167.5 and 175.5\GeV). The choice of the generated mass affects both the extracted signal yield and the signal efficiency. The quadratic interpolation of the measurements with the three different top quark masses is
\ifthenelse{\boolean{cms@external}}{
\begin{multline}\label{eq:interpol}
\frac{\sigma_{\cPqt\cPaqt}(m_\cPqt)}{\sigma_{\cPqt\cPaqt}(m_\cPqt=172.5)}=1.0-2.4\times 10^{-2}\left(m_\cPqt-172.5\right)\\
+8.3\times 10^{-4}\left(m_\cPqt-172.5\right)^2.
\end{multline}
}{
\begin{equation}\label{eq:interpol}
\frac{\sigma_{\cPqt\cPaqt}(m_\cPqt)}{\sigma_{\cPqt\cPaqt}(m_\cPqt=172.5)}=1.0-2.4\times 10^{-2}\left(m_\cPqt-172.5\right)+8.3\times 10^{-4}\left(m_\cPqt-172.5\right)^2.
\end{equation}
}

\subsection{Differential cross sections}\label{sec:diffxsec}

The size of the signal sample allows the differential measurement of the $\cPqt\cPaqt$ production cross section to be performed as a function of various observables. In order to confront the theoretical predictions, the differential cross sections are reported normalized to the inclusive cross section, resulting in a significant cancellation of systematic uncertainties.

\begin{figure*}[hbtp]
  \centering
    \includegraphics[width=0.45\textwidth]{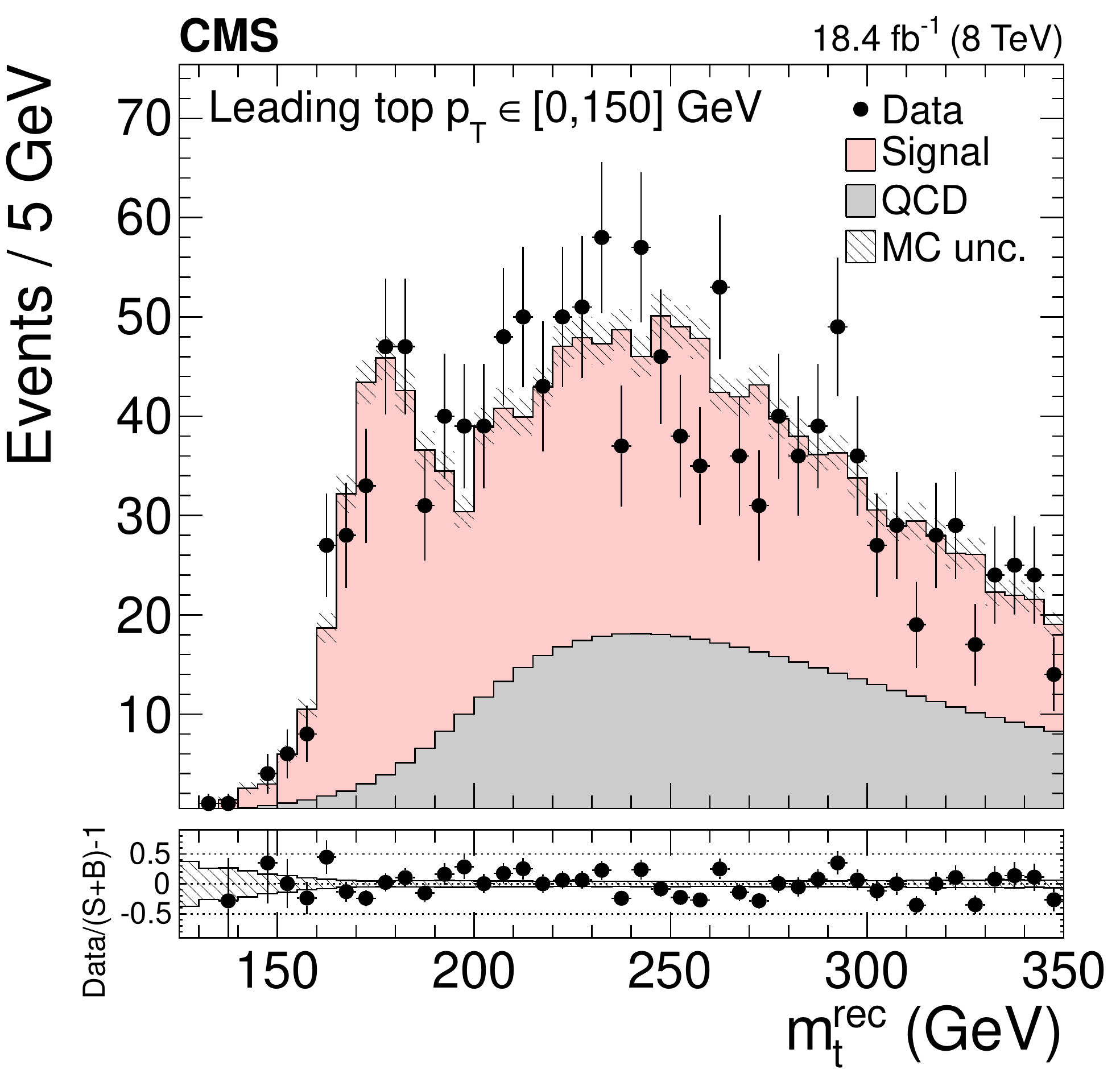}
    \includegraphics[width=0.45\textwidth]{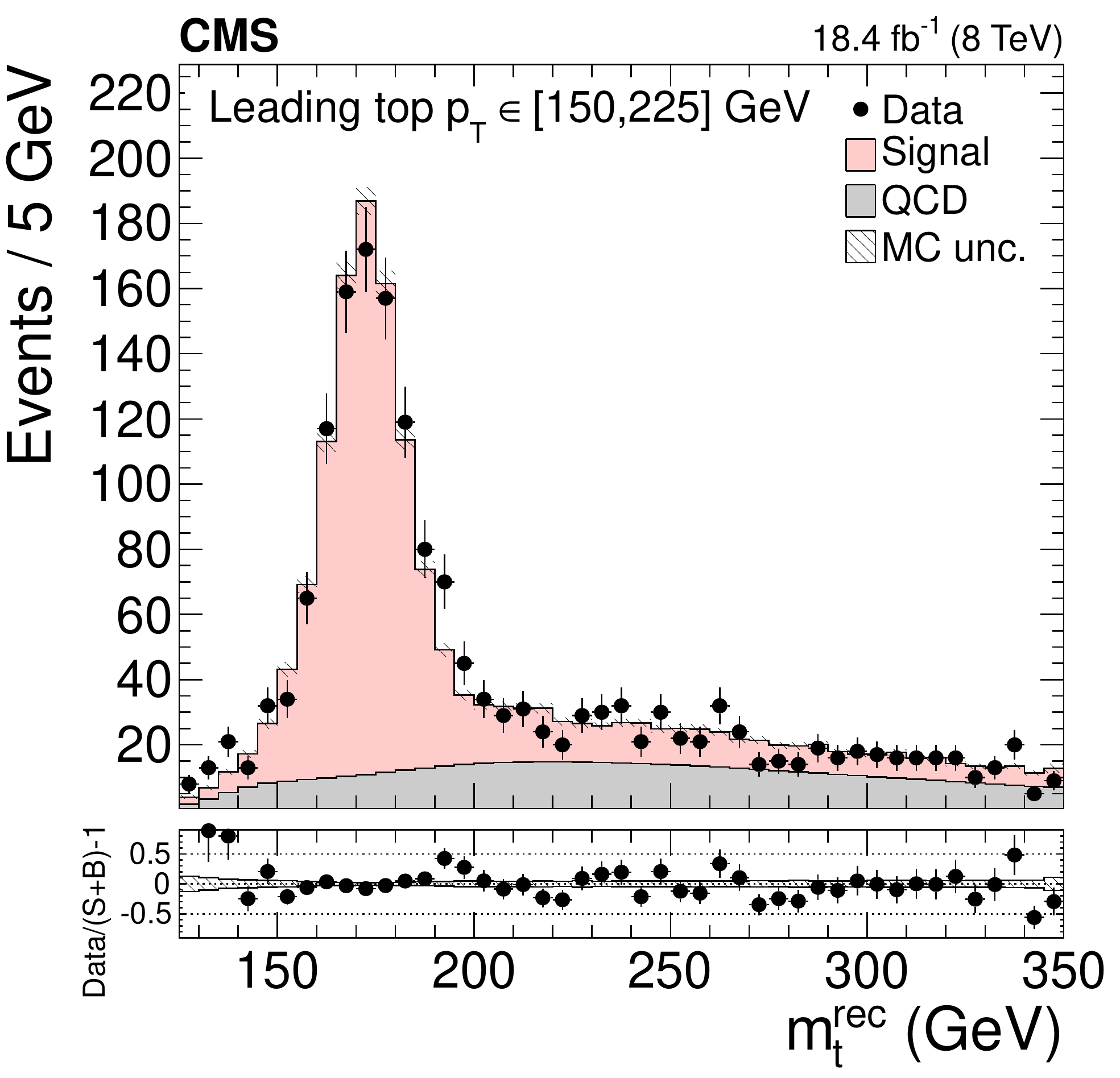}
    \includegraphics[width=0.45\textwidth]{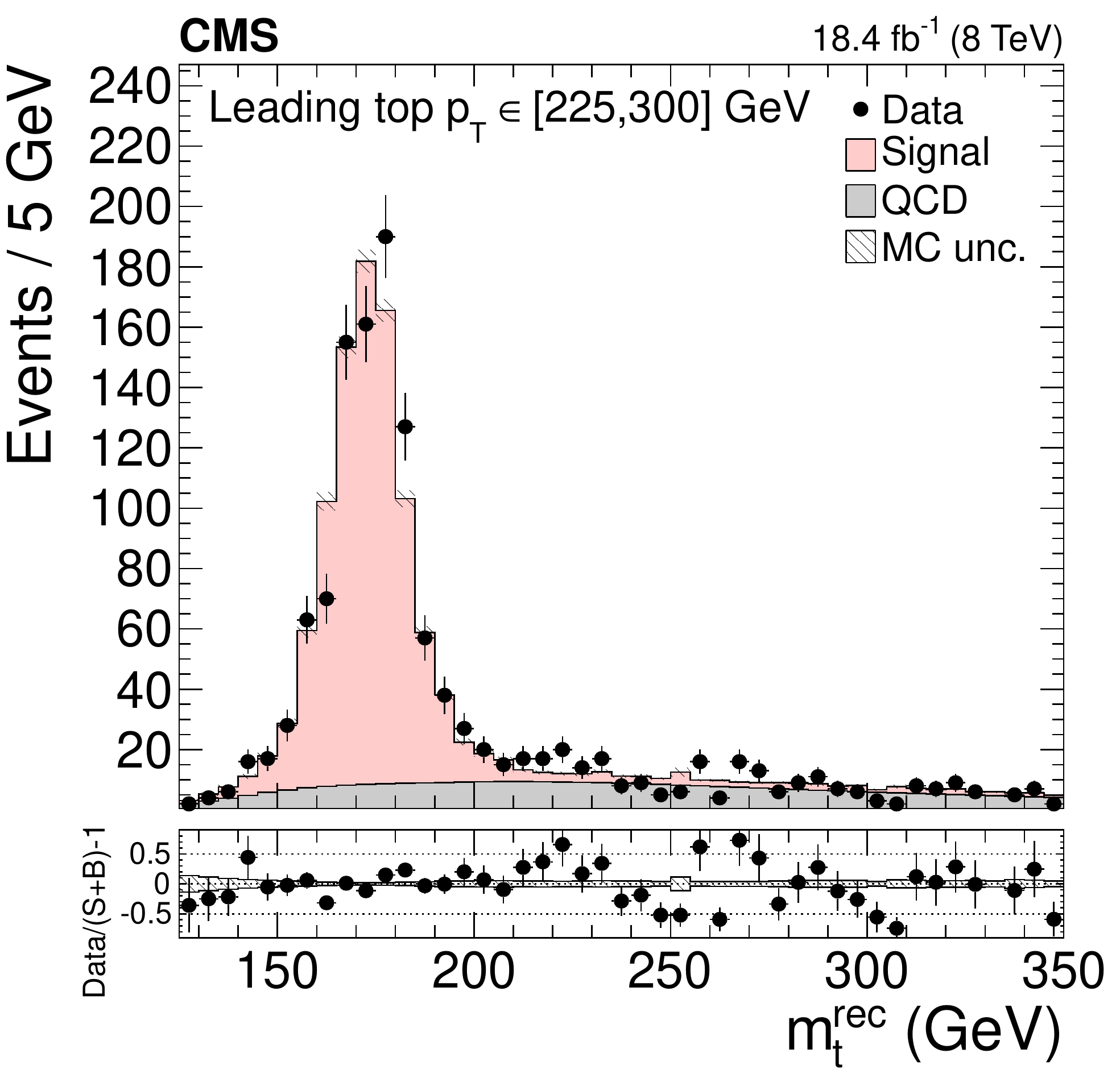}
    \includegraphics[width=0.45\textwidth]{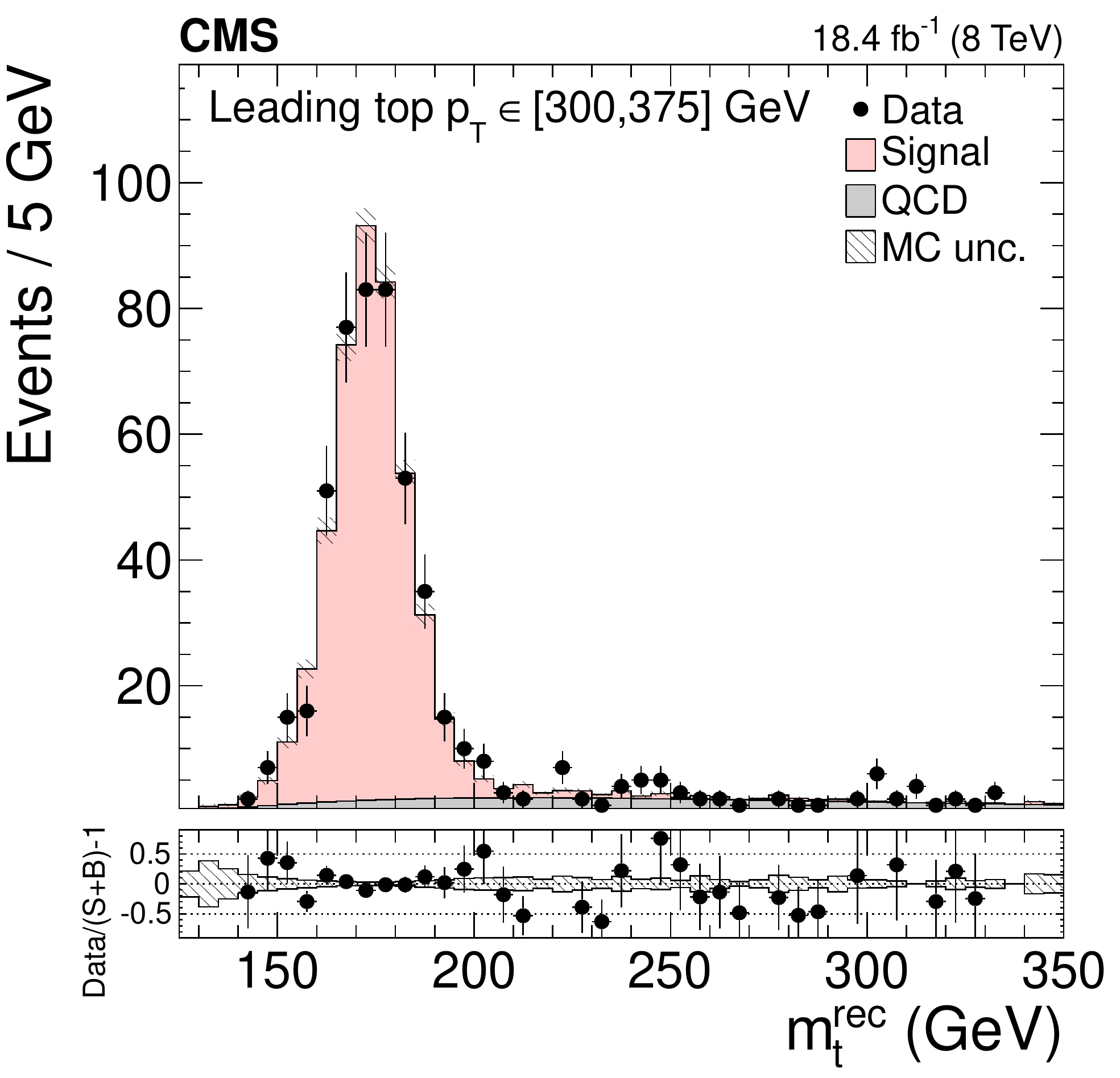}
    \includegraphics[width=0.45\textwidth]{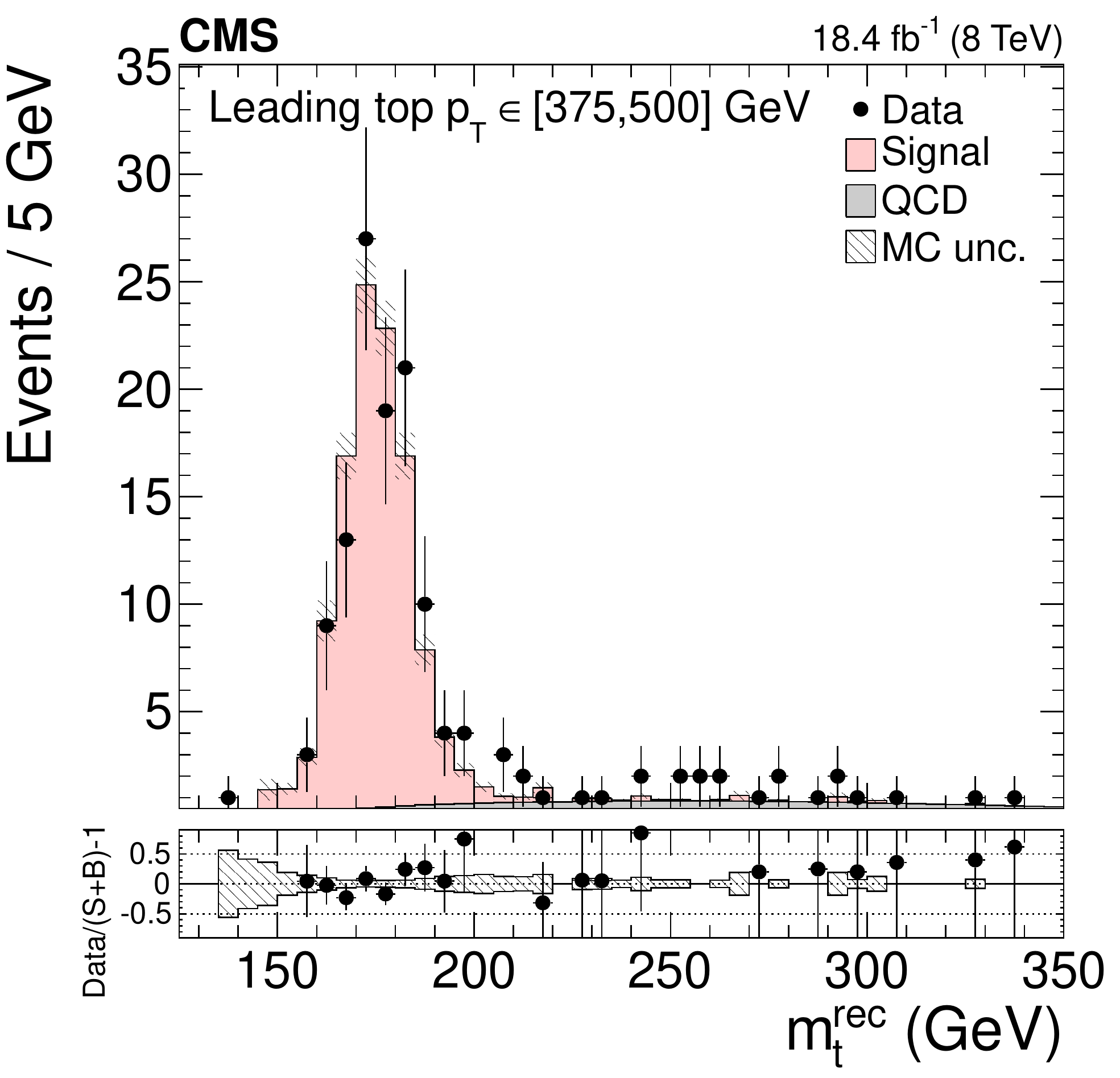}
    \caption{Distribution of the reconstructed top quark mass after the kinematic fit in bins of the leading reconstructed top quark \pt. The normalizations of the $\cPqt\cPaqt$ signal and the QCD multijet background are taken from the template fit to the data. The bottom panels show the fractional difference between the data and the sum of signal and background predictions, with the shaded band representing the MC statistical uncertainty.}
    \label{fig:mtopBins}
  \end{figure*}

The process of measuring the differential cross sections is identical to the inclusive case: in each bin of the observable used to divide the phase space, the signal is extracted from a template fit to the reconstructed top quark mass. Besides the physics interest, the choice of the observables used is mainly motivated by their correlation to $m_{\cPqt}^\text{rec}$, and the ability to extract smooth signal and background templates. The variables chosen are the \pt of the two reconstructed top quarks. Figure~\ref{fig:mtopBins} shows the fitted $m_{\cPqt}^\text{rec}$ distributions in bins of the \pt of the leading top quark.

\begin{figure}[hbtp]
  \centering
    \includegraphics[width=0.49\textwidth]{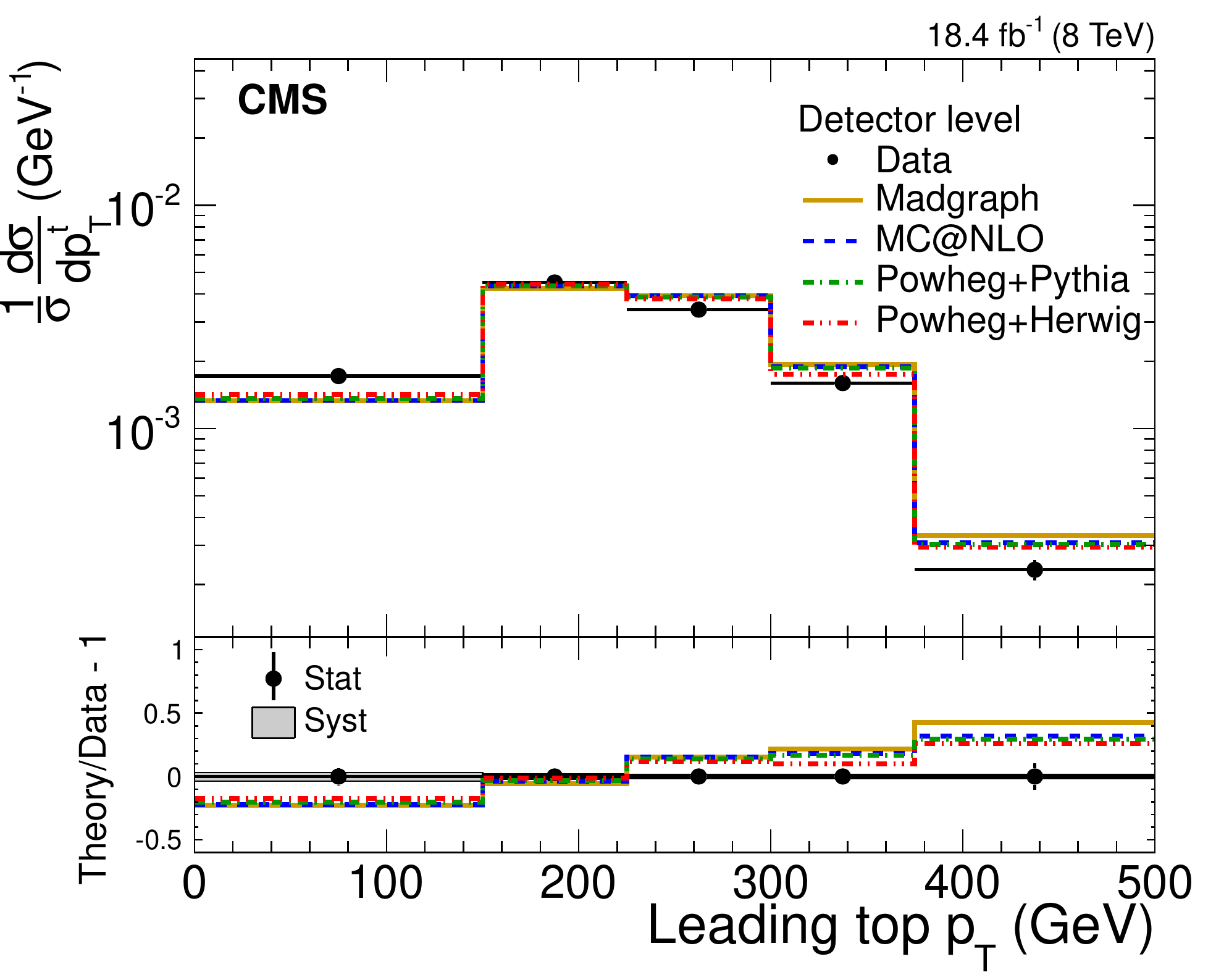}
    \includegraphics[width=0.49\textwidth]{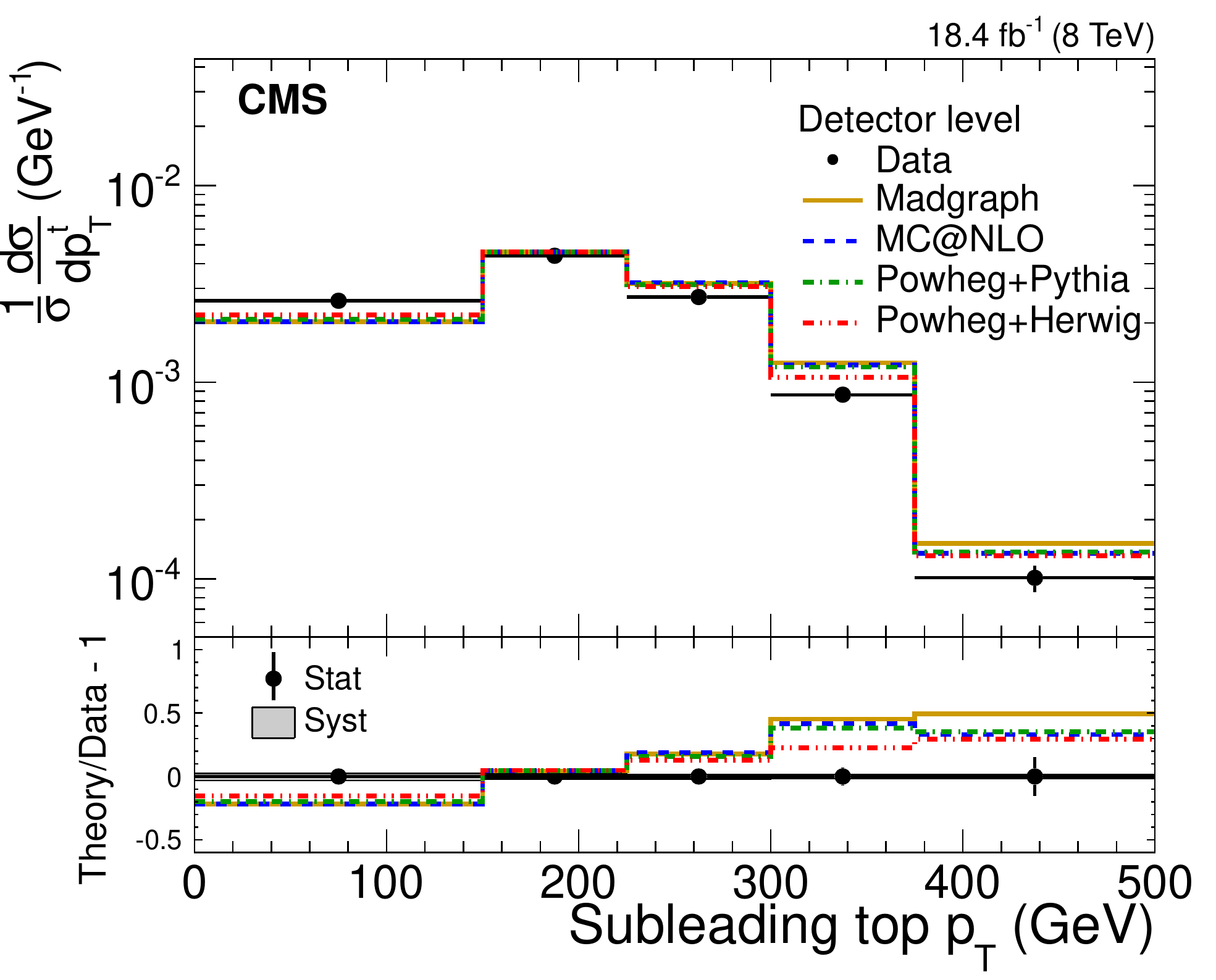}
    \caption{Normalized fiducial differential cross section of the $\cPqt\cPaqt$ production as a function of the leading (\cmsLeft) and subleading (\cmsRight) reconstructed top quark \pt (detector level). The bottom panels show the fractional difference between various MC predictions and the data. Statistical uncertainties are shown with error bars, and systematic uncertainties with the shaded band.}
    \label{fig:diffPtTop}
  \end{figure}

The differential measurements are first reported for the visible fiducial volume, as a function of the reconstructed top \pt (detector level), and then extrapolated to the parton and particle levels. The detector-level result is shown in Fig.~\ref{fig:diffPtTop} and is free of most of the systematic uncertainties affecting the inclusive measurement. The corresponding numerical values are reported in Table~\ref{tab:diffXsecDetector}.

\begin{figure}[hbt]
  \centering
    \includegraphics[width=0.49\textwidth]{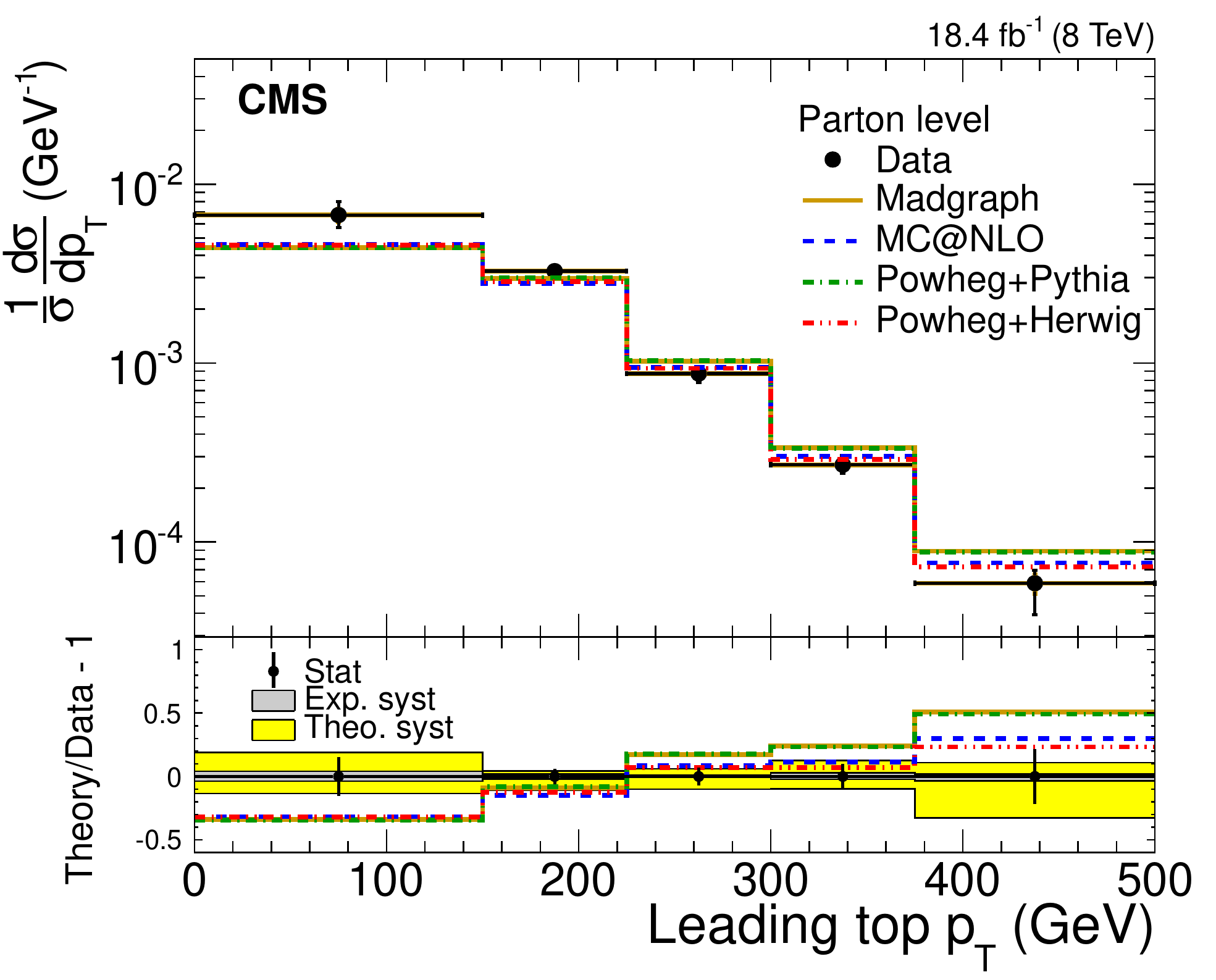}
    \includegraphics[width=0.49\textwidth]{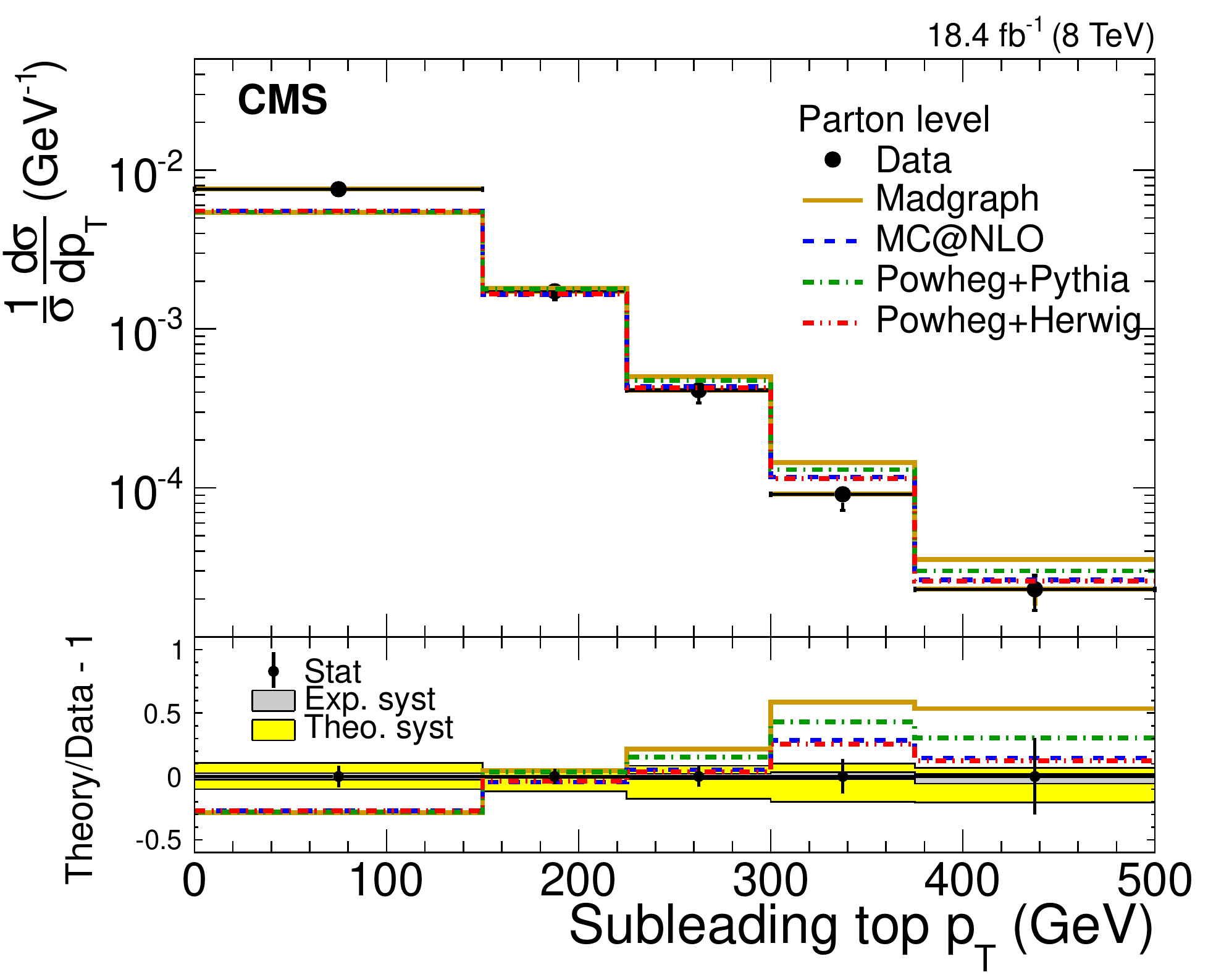}
    \caption{Normalized differential cross section of the $\cPqt\cPaqt$ production at parton level as a function of the leading (\cmsLeft) and subleading (\cmsRight) top quark \pt. The bottom panels show the fractional difference between various MC predictions and the data. Statistical uncertainties are shown with error bars, while theoretical (theo.) and experimental (exp.) systematic uncertainties with the shaded bands.}
    \label{fig:diffXsecParton}
  \end{figure}

The parton-level results shown in Fig.~\ref{fig:diffXsecParton} are obtained from the detector-level measurement, after correcting for bin migration effects and extrapolating to the full phase space using a bin-by-bin acceptance correction. The unfolding of the bin-migration effect is performed with the D'Agostini method~\cite{Bayes}, implemented in the RooUnfold package~\cite{Adye:2011gm}, using the migration matrix derived from the simulation. The uncertainty due to the modeling of the migration matrix and the phase-space extrapolation is estimated by repeating the unfolding and acceptance-correction procedures by varying the systematic sources described in Section~\ref{sec:unc}. The numerical values of the normalized differential cross sections at parton level are reported in Table~\ref{tab:diffXsecParton}. It should be noted that there is a large extrapolation factor involved from the detector-level jets (${\approx}7\times 10^{-4}$ of the signal) to the full parton level, which results in large theoretical uncertainties.

\begin{figure}[hbtp]
  \centering
    \includegraphics[width=0.49\textwidth]{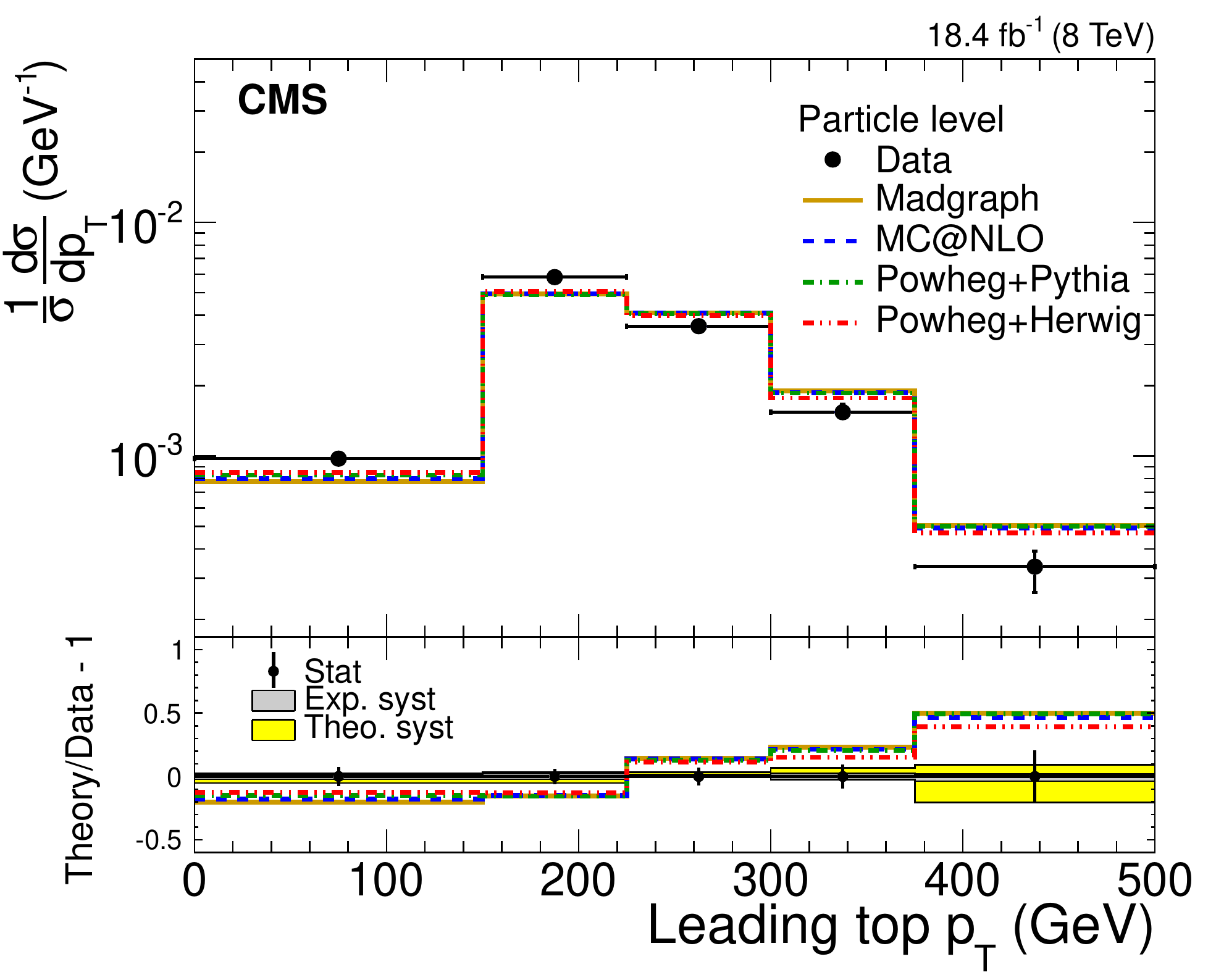}
    \includegraphics[width=0.49\textwidth]{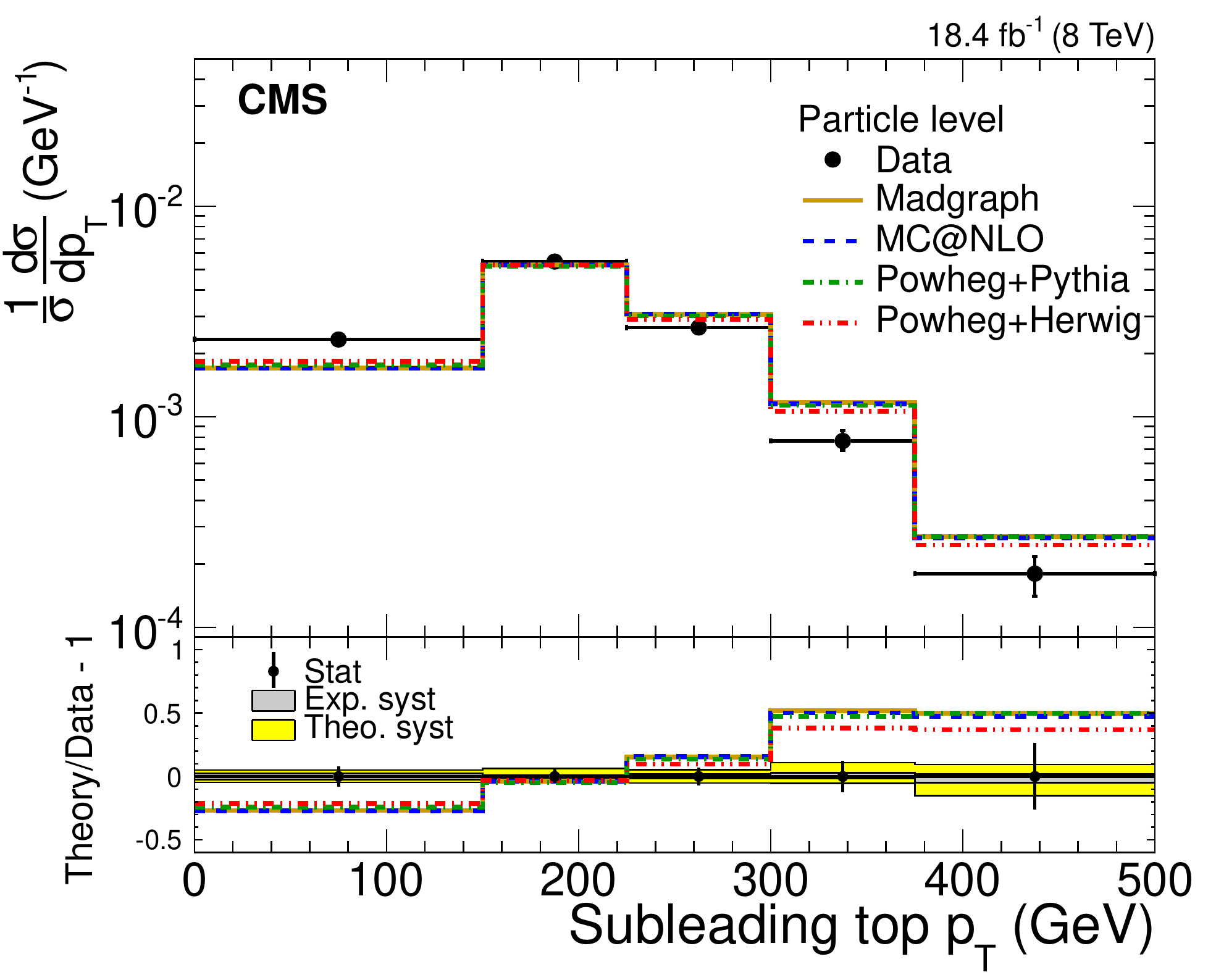}
    \caption{Normalized differential cross section of the $\cPqt\cPaqt$ production at particle level as a function of the leading (\cmsLeft) and subleading (\cmsRight) top quark \pt. The bottom panels show the fractional difference between various MC predictions and the data. Statistical uncertainties are shown with error bars,  while theoretical (theo.) and experimental (exp.) systematic uncertainties with the shaded bands.}
    \label{fig:diffXsecGen}
  \end{figure}

In addition to the parton level, results are reported at particle level, in Fig.~\ref{fig:diffXsecGen}, in a phase space similar to the detector level by construction. This is defined as follows: first, particle jets are built in simulation from all stable particles (including neutrinos) with the same jet clustering algorithm as the detector jets. Then, starting from the six leading jets, the jets associated with B hadrons via matching in $\eta$-$\phi$ ($\Delta R < 0.25$) are identified as the b jet candidates. Events are further selected if $\pt^\text{4th jet} > 60\GeV$ and $\pt^\text{6th jet} > 40\GeV$  and if there are at least two b jets with $\Delta R_{\cPqb\cPqb} > 2.0$. For the selected events, a ``pseudo top quark" is reconstructed from one b jet and the two closest non-b-tagged jets. The particle-level results are obtained in a similar way to the parton level, via unfolding and acceptance correction. The numerical values of the normalized differential cross sections at particle level are reported in Table~\ref{tab:diffXsecParticle}.

The comparison of the measured and predicted differential top quark \pt shapes reveals that the models predict a harder spectrum, both in the leading and in the subleading top quark \pt, in the phase space of the measurement. This effect is also reflected on the jet \pt distributions shown in Fig.~\ref{fig:jetPt}. The \POWHEG+\HERWIG prediction is the closest to the data, but still shows a significant discrepancy. The parton-level results are accompanied by sizeable systematic uncertainties, dominated by the theoretical uncertainties due to the extrapolation to the full phase space. In contrast, the particle-level phase space is much closer to the visible one, and as a result the extrapolation uncertainties are smaller.

\begin{table*}[hbtp]
  \centering
    \topcaption{Normalized differential $\cPqt\cPaqt$ cross section as a function of the \pt of the leading ($\pt^{(1)}$) and subleading ($\pt^{(2)}$) top quarks or antiquarks. The results are presented at detector level in the visible phase space.}
    \label{tab:diffXsecDetector}
    \begin{tabular}{cccc}
      \hline
      \multirow{2}{*}{\pt bin range (\GeVns{})} & \multirow{2}{*}{$\frac{1}{\sigma} \rd\sigma/\rd \pt^{(1)} (\GeVns^{-1})$} & \multirow{2}{*}{stat (\%)} & \multirow{2}{*}{syst (\%)} \\
      & & & \\
      \hline
      $[0,150]$   & $1.72\times10^{-3}$ & $\pm6.7$  & $\pm3.7$ \\
      $[150,225]$ & $4.51\times10^{-3}$ & $\pm3.7$  & $\pm2.0$ \\
      $[225,300]$ & $3.41\times10^{-3}$ & $\pm3.9$  & $\pm1.8$ \\
      $[300,375]$ & $1.60\times10^{-3}$ & $\pm5.3$  & $\pm1.6$ \\
      $[375,500]$ & $2.33\times10^{-4}$ & $\pm10.4$ & $\pm1.7$ \\[2ex]
      \hline
      \multirow{2}{*}{\pt bin range (\GeVns{})} & \multirow{2}{*}{$\frac{1}{\sigma} \rd\sigma/\rd \pt^{(2)} (\GeVns^{-1})$} & \multirow{2}{*}{stat (\%)} & \multirow{2}{*}{syst (\%)} \\
      & & & \\
      \hline
      $[0,150]$   & $2.59\times10^{-3}$ & $\pm3.9$  & $\pm3.3$ \\
      $[150,225]$ & $4.39\times10^{-3}$ & $\pm3.4$  & $\pm1.9$ \\
      $[225,300]$ & $2.71\times10^{-3}$ & $\pm4.1$  & $\pm1.9$ \\
      $[300,375]$ & $8.64\times10^{-4}$ & $\pm7.0$  & $\pm1.8$ \\
      $[375,500]$ & $1.01\times10^{-4}$ & $\pm15.2$ & $\pm1.7$ \\
      \hline
    \end{tabular}
  \end{table*}

\begin{table*}[hbtp]
  \centering
    \topcaption{Normalized differential $\cPqt\cPaqt$ cross section as a function of the \pt of the leading ($\pt^{(1)}$) and subleading ($\pt^{(2)}$) top quarks or antiquarks. The results are presented at parton level in the full phase space.}
    \label{tab:diffXsecParton}
    \begin{tabular}{ccccc}
      \hline
      \multirow{2}{*}{\pt bin range (\GeVns{})} & \multirow{2}{*}{$\frac{1}{\sigma} \rd\sigma/\rd \pt^{(1)} (\GeVns^{-1})$} & \multirow{2}{*}{stat (\%)} & \multirow{2}{*}{exp. syst (\%)} & \multirow{2}{*}{theo. syst (\%)} \\
      & & & \\
      \hline
      $[0,150]$   & $6.72\times10^{-3}$ & $\pm10.8$ & $-3.7, +4.1$  & $-9.7, +14.8$ \\
      $[150,225]$ & $3.27\times10^{-3}$ & $\pm4.3$  & $-2.0, +1.8$  & $-9.0, +2.5$  \\
      $[225,300]$ & $8.73\times10^{-4}$ & $\pm5.0$  & $-0.8, +1.2$  & $-9.3, +4.9$ \\
      $[300,375]$ & $2.70\times10^{-4}$ & $\pm7.1$  & $-2.3, +2.7$  & $-7.5, +9.9$  \\
      $[375,500]$ & $5.88\times10^{-5}$ & $\pm15.2$ & $-3.3, +1.9$  & $-29.4, +9.0$ \\[2ex]
      \hline
      \multirow{2}{*}{\pt bin range (\GeVns{})} & \multirow{2}{*}{$\frac{1}{\sigma} \rd\sigma/\rd \pt^{(2)} (\GeVns^{-1})$} & \multirow{2}{*}{stat (\%)} & \multirow{2}{*}{exp. syst (\%)} & \multirow{2}{*}{theo. syst (\%)} \\
      & & & \\
      \hline
      $[0,150]$   & $7.59\times10^{-3}$ & $\pm6.2$  & $-2.5, +2.7$ & $-7.6, +8.1$  \\
      $[150,225]$ & $1.73\times10^{-3}$ & $\pm4.4$  & $-1.3, +0.7$ & $-10.5, +4.7$ \\
      $[225,300]$ & $4.12\times10^{-4}$ & $\pm5.6$  & $-1.8, +2.2$ & $-15.7, +6.2$ \\
      $[300,375]$ & $9.11\times10^{-5}$ & $\pm9.7$  & $-1.9, +3.3$ & $-18.1, +7.0$ \\
      $[375,500]$ & $2.30\times10^{-5}$ & $\pm21.4$ & $-5.6, +2.0$ & $-15.0, +4.7$ \\
      \hline
    \end{tabular}
  \end{table*}

\begin{table*}[hbtp]
  \centering
    \topcaption{Normalized differential $\cPqt\cPaqt$ cross section as a function of the \pt of the leading ($\pt^{(1)}$) and subleading ($\pt^{(2)}$) top quarks or antiquarks. The results are presented at particle level.}
    \label{tab:diffXsecParticle}
    \begin{tabular}{ccccc}
      \hline
      \multirow{2}{*}{\pt bin range (\GeVns{})} & \multirow{2}{*}{$\frac{1}{\sigma} \rd\sigma/\rd \pt^{(1)} (\GeVns^{-1})$} & \multirow{2}{*}{stat (\%)} & \multirow{2}{*}{exp. syst (\%)} & \multirow{2}{*}{theo. syst (\%)} \\
      & & & \\
      \hline
      $[0,150]$   & $9.75\times10^{-4}$ & $\pm5.2$  & $-2.2, +1.5$ & $-2.9, +1.0$  \\
      $[150,225]$ & $5.83\times10^{-3}$ & $\pm4.4$  & $-2.3, +2.3$ & $-3.0, +1.2$  \\
      $[225,300]$ & $3.60\times10^{-3}$ & $\pm5.0$  & $-0.7, +1.3$ & $-0.0, +3.1$  \\
      $[300,375]$ & $1.54\times10^{-3}$ & $\pm6.8$  & $-2.2, +2.5$ & $-0.4, +4.3$  \\
      $[375,500]$ & $3.36\times10^{-4}$ & $\pm14.6$ & $-3.6, +1.8$ & $-16.9, +7.4$ \\[2ex]
      \hline
      \multirow{2}{*}{\pt bin range (\GeVns{})} & \multirow{2}{*}{$\frac{1}{\sigma} \rd\sigma/\rd \pt^{(2)} (\GeVns^{-1})$} & \multirow{2}{*}{stat (\%)} & \multirow{2}{*}{exp. syst (\%)} & \multirow{2}{*}{theo. syst (\%)} \\
      & & & \\
      \hline
      $[0,150]$   & $2.33\times10^{-3}$ & $\pm5.8$  & $-2.5, +2.5$ & $-3.3, +3.6$ \\
      $[150,225]$ & $5.46\times10^{-3}$ & $\pm3.8$  & $-1.5, +1.2$ & $-3.4, +5.1$ \\
      $[225,300]$ & $2.66\times10^{-3}$ & $\pm5.1$  & $-1.4, +1.8$ & $-3.9, +3.9$ \\
      $[300,375]$ & $7.67\times10^{-4}$ & $\pm8.6$  & $-1.7, +3.0$ & $-4.0, +8.5$ \\
      $[375,500]$ & $1.80\times10^{-4}$ & $\pm18.6$ & $-5.0, +1.9$ & $-11.4, +7.8$ \\
      \hline
    \end{tabular}
  \end{table*}

\section{Summary}

A measurement of the $\cPqt\cPaqt$ production cross section has been performed in the all-jets final state, using pp collision data at $\sqrt{s}=8\TeV$ corresponding to an integrated luminosity of $18.4\fbinv$. The measured inclusive cross section is $275.6\pm 6.1\stat\pm 37.8\syst\pm 7.2\lum$\unit{pb} for a top quark mass of $172.5\GeV$, in agreement with the standard model prediction of $252.9^{+6.4}_{-8.6}\,(\text{scale})\pm11.7\,(\text{PDF}+\alpha_S)$\unit{pb} as calculated with the \textsc{Top++} (v. 2.0) program~\cite{Czakon:2011xx} at next-to-next-to-leading order in perturbative QCD, including soft-gluon resummation at next-to-next-to-leading-log order~\cite{Czakon:2013goa}, and assuming a top-quark mass $m_{\cPqt}=172.5\GeV$. Also reported are the fiducial normalized differential cross sections as a function of the leading and subleading top quark \pt. Compared to QCD predictions, the measurement shows a significantly softer top quark \pt spectrum. The differential cross sections are also extrapolated to the full partonic phase space, as well as to particle level, and can be used to tune Monte Carlo models.

\begin{acknowledgments}
We congratulate our colleagues in the CERN accelerator departments for the excellent performance of the LHC and thank the technical and administrative staffs at CERN and at other CMS institutes for their contributions to the success of the CMS effort. In addition, we gratefully acknowledge the computing centres and personnel of the Worldwide LHC Computing Grid for delivering so effectively the computing infrastructure essential to our analyses. Finally, we acknowledge the enduring support for the construction and operation of the LHC and the CMS detector provided by the following funding agencies: BMWFW and FWF (Austria); FNRS and FWO (Belgium); CNPq, CAPES, FAPERJ, and FAPESP (Brazil); MES (Bulgaria); CERN; CAS, MoST, and NSFC (China); COLCIENCIAS (Colombia); MSES and CSF (Croatia); RPF (Cyprus); MoER, ERC IUT and ERDF (Estonia); Academy of Finland, MEC, and HIP (Finland); CEA and CNRS/IN2P3 (France); BMBF, DFG, and HGF (Germany); GSRT (Greece); OTKA and NIH (Hungary); DAE and DST (India); IPM (Iran); SFI (Ireland); INFN (Italy); MSIP and NRF (Republic of Korea); LAS (Lithuania); MOE and UM (Malaysia); CINVESTAV, CONACYT, SEP, and UASLP-FAI (Mexico); MBIE (New Zealand); PAEC (Pakistan); MSHE and NSC (Poland); FCT (Portugal); JINR (Dubna); MON, RosAtom, RAS and RFBR (Russia); MESTD (Serbia); SEIDI and CPAN (Spain); Swiss Funding Agencies (Switzerland); MST (Taipei); ThEPCenter, IPST, STAR and NSTDA (Thailand); TUBITAK and TAEK (Turkey); NASU and SFFR (Ukraine); STFC (United Kingdom); DOE and NSF (USA).

Individuals have received support from the Marie-Curie programme and the European Research Council and EPLANET (European Union); the Leventis Foundation; the A. P. Sloan Foundation; the Alexander von Humboldt Foundation; the Belgian Federal Science Policy Office; the Fonds pour la Formation \`a la Recherche dans l'Industrie et dans l'Agriculture (FRIA-Belgium); the Agentschap voor Innovatie door Wetenschap en Technologie (IWT-Belgium); the Ministry of Education, Youth and Sports (MEYS) of the Czech Republic; the Council of Science and Industrial Research, India; the HOMING PLUS programme of the Foundation for Polish Science, cofinanced from European Union, Regional Development Fund; the OPUS programme of the National Science Center (Poland); the Compagnia di San Paolo (Torino); the Consorzio per la Fisica (Trieste); MIUR project 20108T4XTM (Italy); the Thalis and Aristeia programmes cofinanced by EU-ESF and the Greek NSRF; the National Priorities Research Program by Qatar National Research Fund; the Rachadapisek Sompot Fund for Postdoctoral Fellowship, Chulalongkorn University (Thailand); and the Welch Foundation, contract C-1845.
\end{acknowledgments}

\clearpage
\bibliography{auto_generated}

\cleardoublepage \appendix\section{The CMS Collaboration \label{app:collab}}\begin{sloppypar}\hyphenpenalty=5000\widowpenalty=500\clubpenalty=5000\input{TOP-14-018-authorlist.tex}\end{sloppypar}
\end{document}

%% file: TOP-14-018-authorlist.tex
\textbf{Yerevan Physics Institute,  Yerevan,  Armenia}\\*[0pt]
V.~Khachatryan, A.M.~Sirunyan, A.~Tumasyan
\vskip\cmsinstskip
\textbf{Institut f\"{u}r Hochenergiephysik der OeAW,  Wien,  Austria}\\*[0pt]
W.~Adam, E.~Asilar, T.~Bergauer, J.~Brandstetter, E.~Brondolin, M.~Dragicevic, J.~Er\"{o}, M.~Flechl, M.~Friedl, R.~Fr\"{u}hwirth\cmsAuthorMark{1}, V.M.~Ghete, C.~Hartl, N.~H\"{o}rmann, J.~Hrubec, M.~Jeitler\cmsAuthorMark{1}, V.~Kn\"{u}nz, A.~K\"{o}nig, M.~Krammer\cmsAuthorMark{1}, I.~Kr\"{a}tschmer, D.~Liko, T.~Matsushita, I.~Mikulec, D.~Rabady\cmsAuthorMark{2}, B.~Rahbaran, H.~Rohringer, J.~Schieck\cmsAuthorMark{1}, R.~Sch\"{o}fbeck, J.~Strauss, W.~Treberer-Treberspurg, W.~Waltenberger, C.-E.~Wulz\cmsAuthorMark{1}
% ...
\vskip\cmsinstskip
% ...
1:~~Also at Vienna University of Technology, Vienna, Austria\\
2:~~Also at CERN, European Organization for Nuclear Research, Geneva, Switzerland\\
% ...


%% file: TOP-14-018_temp.bbl
\providecommand{\href}[2]{#2}\begingroup\raggedright\begin{thebibliography}{10}%
\makeatletter
\providecommand{\hrefCMSnoop }[0]{\@secondoftwo}%
\makeatother
\providecommand{\doi}{\texttt{doi:}\begingroup \urlstyle{tt}\Url}

\bibitem{Aad:2010ey}
\hrefCMSnoop {}{{ATLAS Collaboration}, ``{Measurement of the top quark-pair
  production cross section with ATLAS in pp collisions at $\sqrt{s}=7$ TeV}'',}
  \textit{ Eur. Phys. J. C} \textbf{ 71} (2011) 1577,
  \href{http://dx.doi.org/10.1140/epjc/s10052-011-1577-6}{\doi{10.1140/epjc/s10052-011-1577-6}},
\href{http://www.arXiv.org/abs/1012.1792}{\texttt{arXiv:1012.1792}}.

\bibitem{Aad:2011yb}
\hrefCMSnoop {}{{ATLAS Collaboration}, ``{Measurement of the top quark pair
  production cross section in $pp$ collisions at $\sqrt{s}=7$ TeV in dilepton
  final states with ATLAS}'',} \textit{ Phys. Lett. B} \textbf{ 707} (2012)
  459,
  \href{http://dx.doi.org/10.1016/j.physletb.2011.12.055}{\doi{10.1016/j.physletb.2011.12.055}},
\href{http://www.arXiv.org/abs/1108.3699}{\texttt{arXiv:1108.3699}}.

\bibitem{Aad:2012qf}
\hrefCMSnoop {}{{ATLAS Collaboration}, ``{Measurement of the top quark pair
  production cross-section with ATLAS in the single lepton channel}'',}
  \textit{ Phys. Lett. B} \textbf{ 711} (2012) 244,
  \href{http://dx.doi.org/10.1016/j.physletb.2012.03.083}{\doi{10.1016/j.physletb.2012.03.083}},
\href{http://www.arXiv.org/abs/1201.1889}{\texttt{arXiv:1201.1889}}.

\bibitem{ATLAS:2012aa}
\hrefCMSnoop {}{{ATLAS Collaboration}, ``{Measurement of the cross section for
  top-quark pair production in $pp$ collisions at $\sqrt{s}=7$ TeV with the
  ATLAS detector using final states with two high-pt leptons}'',} \textit{
  JHEP} \textbf{ 05} (2012) 059,
  \href{http://dx.doi.org/10.1007/JHEP05(2012)059}{\doi{10.1007/JHEP05(2012)059}},
\href{http://www.arXiv.org/abs/1202.4892}{\texttt{arXiv:1202.4892}}.

\bibitem{Aad:2012mza}
\hrefCMSnoop {}{{ATLAS Collaboration}, ``{Measurement of the top quark pair
  cross section with ATLAS in pp collisions at $\sqrt{s}=7$ TeV using final
  states with an electron or a muon and a hadronically decaying $\tau$
  lepton}'',} \textit{ Phys. Lett. B} \textbf{ 717} (2012) 89,
  \href{http://dx.doi.org/10.1016/j.physletb.2012.09.032}{\doi{10.1016/j.physletb.2012.09.032}},
\href{http://www.arXiv.org/abs/1205.2067}{\texttt{arXiv:1205.2067}}.

\bibitem{Aad:2012vip}
\hrefCMSnoop {}{{ATLAS Collaboration}, ``{Measurement of the ttbar production
  cross section in the tau+jets channel using the ATLAS detector}'',} \textit{
  Eur. Phys. J. C} \textbf{ 73} (2013) 2328,
  \href{http://dx.doi.org/10.1140/epjc/s10052-013-2328-7}{\doi{10.1140/epjc/s10052-013-2328-7}},
\href{http://www.arXiv.org/abs/1211.7205}{\texttt{arXiv:1211.7205}}.

\bibitem{Aad:2014kva}
\hrefCMSnoop {}{{ATLAS Collaboration}, ``{Measurement of the $t\overline{t}$
  production cross-section using $e\mu $ events with $b$ -tagged jets in $pp$
  collisions at $\sqrt{s}=7$ and 8 TeV with the ATLAS detector}'',} \textit{
  Eur. Phys. J. C} \textbf{ 74} (2014) 3109,
  \href{http://dx.doi.org/10.1140/epjc/s10052-014-3109-7}{\doi{10.1140/epjc/s10052-014-3109-7}},
\href{http://www.arXiv.org/abs/1406.5375}{\texttt{arXiv:1406.5375}}.

\bibitem{Aad:2014zka}
\hrefCMSnoop {}{{ATLAS Collaboration}, ``{Measurements of normalized
  differential cross sections for $t\bar{t}$ production in pp collisions at
  $\sqrt{s}=7$~TeV using the ATLAS detector}'',} \textit{ Phys. Rev. D}
  \textbf{ 90} (2014) 072004,
  \href{http://dx.doi.org/10.1103/PhysRevD.90.072004}{\doi{10.1103/PhysRevD.90.072004}},
\href{http://www.arXiv.org/abs/1407.0371}{\texttt{arXiv:1407.0371}}.

\bibitem{Aad:2015eia}
\hrefCMSnoop {}{{ATLAS Collaboration}, ``{Differential top-antitop
  cross-section measurements as a function of observables constructed from
  final-state particles using pp collisions at $\sqrt{s}=7$ TeV in the ATLAS
  detector}'',} \textit{ JHEP} \textbf{ 06} (2015) 100,
  \href{http://dx.doi.org/10.1007/JHEP06(2015)100}{\doi{10.1007/JHEP06(2015)100}},
\href{http://www.arXiv.org/abs/1502.05923}{\texttt{arXiv:1502.05923}}.

\bibitem{Aad:2015pga}
\hrefCMSnoop {}{{ATLAS Collaboration}, ``{Measurement of the top pair
  production cross section in 8 TeV proton-proton collisions using kinematic
  information in the lepton+jets final state with ATLAS}'',} \textit{ Phys.
  Rev. D} \textbf{ 91} (2015) 112013,
  \href{http://dx.doi.org/10.1103/PhysRevD.91.112013}{\doi{10.1103/PhysRevD.91.112013}},
\href{http://www.arXiv.org/abs/1504.04251}{\texttt{arXiv:1504.04251}}.

\bibitem{Aad:2015dya}
\hrefCMSnoop {}{{ATLAS Collaboration}, ``{Measurement of the top quark
  branching ratios into channels with leptons and quarks with the ATLAS
  detector}'',} \textit{ Phys. Rev. D} \textbf{ 92} (2015) 072005,
  \href{http://dx.doi.org/10.1103/PhysRevD.92.072005}{\doi{10.1103/PhysRevD.92.072005}},
\href{http://www.arXiv.org/abs/1506.05074}{\texttt{arXiv:1506.05074}}.

\bibitem{Khachatryan:2010ez}
\hrefCMSnoop {}{{CMS Collaboration}, ``{First measurement of the cross section
  for top-quark pair production in proton-proton collisions at
  $\sqrt{s}=7$\TeV}'',} \textit{ Phys. Lett. B} \textbf{ 695} (2011) 424,
  \href{http://dx.doi.org/10.1016/j.physletb.2010.11.058}{\doi{10.1016/j.physletb.2010.11.058}},
\href{http://www.arXiv.org/abs/1010.5994}{\texttt{arXiv:1010.5994}}.

\bibitem{Chatrchyan:2011nb}
\hrefCMSnoop {}{{CMS Collaboration}, ``{Measurement of the $\ttbar$ production
  cross section and the top quark mass in the dilepton channel in pp collisions
  at $\sqrt{s}=7$ TeV}'',} \textit{ JHEP} \textbf{ 07} (2011) 049,
  \href{http://dx.doi.org/10.1007/JHEP07(2011)049}{\doi{10.1007/JHEP07(2011)049}},
\href{http://www.arXiv.org/abs/1105.5661}{\texttt{arXiv:1105.5661}}.

\bibitem{Chatrchyan:2011ew}
\hrefCMSnoop {}{{CMS Collaboration}, ``{Measurement of the $\ttbar$ production
  cross section in pp collisions at $\sqrt{s}=7$ TeV using the kinematic
  properties of events with leptons and jets}'',} \textit{ Eur. Phys. J. C}
  \textbf{ 71} (2011) 1721,
  \href{http://dx.doi.org/10.1140/epjc/s10052-011-1721-3}{\doi{10.1140/epjc/s10052-011-1721-3}},
\href{http://www.arXiv.org/abs/1106.0902}{\texttt{arXiv:1106.0902}}.

\bibitem{Chatrchyan:2011yy}
\hrefCMSnoop {}{{CMS Collaboration}, ``{Measurement of the $t \bar{t}$
  Production Cross Section in $pp$ Collisions at 7 TeV in Lepton + Jets Events
  Using $b$-quark Jet Identification}'',} \textit{ Phys. Rev. D} \textbf{ 84}
  (2011) 092004,
  \href{http://dx.doi.org/10.1103/PhysRevD.84.092004}{\doi{10.1103/PhysRevD.84.092004}},
\href{http://www.arXiv.org/abs/1108.3773}{\texttt{arXiv:1108.3773}}.

\bibitem{Chatrchyan:2012vs}
\hrefCMSnoop {}{{CMS Collaboration}, ``{Measurement of the $\ttbar$ production
  cross section in pp collisions at $\sqrt{s}=7$ TeV in dilepton final states
  containing a $\tau$}'',} \textit{ Phys. Rev. D} \textbf{ 85} (2012) 112007,
  \href{http://dx.doi.org/10.1103/PhysRevD.85.112007}{\doi{10.1103/PhysRevD.85.112007}},
\href{http://www.arXiv.org/abs/1203.6810}{\texttt{arXiv:1203.6810}}.

\bibitem{Chatrchyan:2012bra}
\hrefCMSnoop {}{{CMS Collaboration}, ``{Measurement of the $\ttbar$ production
  cross section in the dilepton channel in pp collisions at $\sqrt{s}=7$
  TeV}'',} \textit{ JHEP} \textbf{ 11} (2012) 067,
  \href{http://dx.doi.org/10.1007/JHEP11(2012)067}{\doi{10.1007/JHEP11(2012)067}},
\href{http://www.arXiv.org/abs/1208.2671}{\texttt{arXiv:1208.2671}}.

\bibitem{Chatrchyan:2012saa}
\hrefCMSnoop {}{{CMS Collaboration}, ``{Measurement of differential top-quark
  pair production cross sections in $pp$ colisions at $\sqrt{s}=7$ TeV}'',}
  \textit{ Eur. Phys. J. C} \textbf{ 73} (2013) 2339,
  \href{http://dx.doi.org/10.1140/epjc/s10052-013-2339-4}{\doi{10.1140/epjc/s10052-013-2339-4}},
\href{http://www.arXiv.org/abs/1211.2220}{\texttt{arXiv:1211.2220}}.

\bibitem{Chatrchyan:2012ria}
\hrefCMSnoop {}{{CMS Collaboration}, ``{Measurement of the $\ttbar$ production
  cross section in pp collisions at $\sqrt{s}=7$\TeV with lepton+jets final
  states}'',} \textit{ Phys. Lett. B} \textbf{ 720} (2013) 83,
  \href{http://dx.doi.org/10.1016/j.physletb.2013.02.021}{\doi{10.1016/j.physletb.2013.02.021}},
\href{http://www.arXiv.org/abs/1212.6682}{\texttt{arXiv:1212.6682}}.

\bibitem{Chatrchyan:2013kff}
\hrefCMSnoop {}{{CMS Collaboration}, ``{Measurement of the $\ttbar$ production
  cross section in the tau+jets channel in pp collisions at
  $\sqrt{s}=7$\TeV}'',} \textit{ Eur. Phys. J. C} \textbf{ 73} (2013) 2386,
  \href{http://dx.doi.org/10.1140/epjc/s10052-013-2386-x}{\doi{10.1140/epjc/s10052-013-2386-x}},
\href{http://www.arXiv.org/abs/1301.5755}{\texttt{arXiv:1301.5755}}.

\bibitem{Chatrchyan:2013ual}
\hrefCMSnoop {}{{CMS Collaboration}, ``{Measurement of the $t\bar{t}$
  production cross section in the all-jet final state in pp collisions at
  $\sqrt{s}=7$ TeV}'',} \textit{ JHEP} \textbf{ 05} (2013) 065,
  \href{http://dx.doi.org/10.1007/JHEP05(2013)065}{\doi{10.1007/JHEP05(2013)065}},
\href{http://www.arXiv.org/abs/1302.0508}{\texttt{arXiv:1302.0508}}.

\bibitem{Chatrchyan:2013faa}
\hrefCMSnoop {}{{CMS Collaboration}, ``{Measurement of the $t \bar{t}$
  production cross section in the dilepton channel in pp collisions at
  $\sqrt{s} = 8$ TeV}'',} \textit{ JHEP} \textbf{ 02} (2014) 024,
  \href{http://dx.doi.org/10.1007/JHEP02(2014)024}{\doi{10.1007/JHEP02(2014)024}},
  \href{http://www.arXiv.org/abs/1312.7582}{\texttt{arXiv:1312.7582}}.
[Erratum: \DOI{10.1007/JHEP02(2014)102}].

\bibitem{Khachatryan:2014loa}
\hrefCMSnoop {}{{CMS Collaboration}, ``{Measurement of the $\ttbar$ production
  cross section in pp collisions at $\sqrt{s}=8$ TeV in dilepton final states
  containing one $\tau$ lepton}'',} \textit{ Phys. Lett. B} \textbf{ 739}
  (2014) 23,
  \href{http://dx.doi.org/10.1016/j.physletb.2014.10.032}{\doi{10.1016/j.physletb.2014.10.032}},
\href{http://www.arXiv.org/abs/1407.6643}{\texttt{arXiv:1407.6643}}.

\bibitem{Khachatryan:2015oqa}
\hrefCMSnoop {}{{CMS Collaboration}, ``{Measurement of the differential cross
  section for top quark pair production in pp collisions at $\sqrt{s}=8$
  TeV}'',} \textit{ Eur. Phys. J. C} \textbf{ 75} (2015) 542,
  \href{http://dx.doi.org/10.1140/epjc/s10052-015-3709-x}{\doi{10.1140/epjc/s10052-015-3709-x}},
\href{http://www.arXiv.org/abs/1505.04480}{\texttt{arXiv:1505.04480}}.

\bibitem{Aaltonen:2013wca}
\hrefCMSnoop {}{{CDF, D0} Collaboration, ``{Combination of measurements of the
  top-quark pair production cross section from the Tevatron Collider}'',}
  \textit{ Phys. Rev. D} \textbf{ 89} (2014) 072001,
  \href{http://dx.doi.org/10.1103/PhysRevD.89.072001}{\doi{10.1103/PhysRevD.89.072001}},
\href{http://www.arXiv.org/abs/1309.7570}{\texttt{arXiv:1309.7570}}.

\bibitem{Aaltonen:2010pe}
\hrefCMSnoop {}{{CDF} Collaboration, ``{Measurement of the Top Quark Mass and
  $p\bar{p}\rightarrow$ $t\bar{t}$ Cross Section in the All-Hadronic Mode with
  the CDFII Detector}'',} \textit{ Phys. Rev. D} \textbf{ 81} (2010) 052011,
  \href{http://dx.doi.org/10.1103/PhysRevD.81.052011}{\doi{10.1103/PhysRevD.81.052011}},
\href{http://www.arXiv.org/abs/1002.0365}{\texttt{arXiv:1002.0365}}.

\bibitem{Abazov:2009ss}
\hrefCMSnoop {}{{D0} Collaboration, ``{Measurement of the $t\bar{t}$ cross
  section using high-multiplicity jet events}'',} \textit{ Phys. Rev. D}
  \textbf{ 82} (2010) 032002,
  \href{http://dx.doi.org/10.1103/PhysRevD.82.032002}{\doi{10.1103/PhysRevD.82.032002}},
\href{http://www.arXiv.org/abs/0911.4286}{\texttt{arXiv:0911.4286}}.

\bibitem{Chatrchyan:2008zzk}
\hrefCMSnoop {}{{CMS Collaboration}, ``The {CMS} experiment at the {CERN}
  {LHC}'',} \textit{ JINST} \textbf{ 3} (2008) S08004,
  \href{http://dx.doi.org/10.1088/1748-0221/3/08/S08004}{\doi{10.1088/1748-0221/3/08/S08004}}.

\bibitem{Alwall:2014hca}
J.~Alwall\hrefCMSnoop {}{ {et~al.}, ``{The automated computation of tree-level
  and next-to-leading order differential cross sections, and their matching to
  parton shower simulations}'',} \textit{ JHEP} \textbf{ 07} (2014) 079,
  \href{http://dx.doi.org/10.1007/JHEP07(2014)079}{\doi{10.1007/JHEP07(2014)079}},
\href{http://www.arXiv.org/abs/1405.0301}{\texttt{arXiv:1405.0301}}.

\bibitem{Artoisenet:2012st}
\hrefCMSnoop {}{P.~Artoisenet, R.~Frederix, O.~Mattelaer, and R.~Rietkerk,
  ``{Automatic spin-entangled decays of heavy resonances in Monte Carlo
  simulations}'',} \textit{ JHEP} \textbf{ 03} (2013) 015,
  \href{http://dx.doi.org/10.1007/JHEP03(2013)015}{\doi{10.1007/JHEP03(2013)015}},
\href{http://www.arXiv.org/abs/1212.3460}{\texttt{arXiv:1212.3460}}.

\bibitem{cteq}
J.~Pumplin\hrefCMSnoop {}{ {et~al.}, ``{New generation of parton distributions
  with uncertainties from global QCD analysis}'',} \textit{ JHEP} \textbf{ 07}
  (2002) 012,
  \href{http://dx.doi.org/10.1088/1126-6708/2002/07/012}{\doi{10.1088/1126-6708/2002/07/012}},
\href{http://www.arXiv.org/abs/hep-ph/0201195}{\texttt{arXiv:hep-ph/0201195}}.

\bibitem{Sjostrand:2006za}
\hrefCMSnoop {}{T.~Sj{\"o}strand, S.~Mrenna, and P.~Z. Skands, ``{PYTHIA} 6.4
  physics and manual'',} \textit{ JHEP} \textbf{ 05} (2006) 026,
  \href{http://dx.doi.org/10.1088/1126-6708/2006/05/026}{\doi{10.1088/1126-6708/2006/05/026}},
\href{http://www.arXiv.org/abs/hep-ph/0603175}{\texttt{arXiv:hep-ph/0603175}}.

\bibitem{MLM}
\hrefCMSnoop {}{M.~L. Mangano, M.~Moretti, F.~Piccinini, and M.~Treccani,
  ``Matching matrix elements and shower evolution for top-quark production in
  hadronic collisions'',} \textit{ JHEP} \textbf{ 01} (2007) 013,
  \href{http://dx.doi.org/10.1088/1126-6708/2007/01/013}{\doi{10.1088/1126-6708/2007/01/013}},
  \href{http://www.arXiv.org/abs/hep-ex/0611129}{\texttt{arXiv:hep-ex/0611129}}.

\bibitem{Field:2010bc}
\href {http://inspirehep.net/record/873443/files/arXiv:1010.3558.pdf}{R.~Field,
  ``{Early LHC Underlying Event Data - Findings and Surprises}'',} in \textit{
  {Hadron collider physics. Proceedings, 22nd Conference, HCP 2010, Toronto,
  Canada, August 23-27, 2010}}.
\newblock 2010.
\newblock
\href{http://www.arXiv.org/abs/1010.3558}{\texttt{arXiv:1010.3558}}.
\newblock

\bibitem{geant}
\hrefCMSnoop {}{{GEANT4} Collaboration, ``{GEANT4}---a simulation toolkit'',}
  \textit{ Nucl. Instrum. Meth. A} \textbf{ 506} (2003) 250,
\href{http://dx.doi.org/10.1016/S0168-9002(03)01368-8}{\doi{10.1016/S0168-9002(03)01368-8}}.

\bibitem{mcatnlo}
\hrefCMSnoop {}{S.~Frixione and B.~R. Webber, ``{Matching NLO QCD computations
  and parton shower simulations}'',} \textit{ JHEP} \textbf{ 06} (2002) 29,
  \href{http://dx.doi.org/10.1088/1126-6708/2002/06/029}{\doi{10.1088/1126-6708/2002/06/029}},
  \href{http://www.arXiv.org/abs/hep-ph/0204244}{\texttt{arXiv:hep-ph/0204244}}.

\bibitem{Nason:2004rx}
\hrefCMSnoop {}{P.~Nason, ``{A New method for combining NLO QCD with shower
  Monte Carlo algorithms}'',} \textit{ JHEP} \textbf{ 11} (2004) 040,
  \href{http://dx.doi.org/10.1088/1126-6708/2004/11/040}{\doi{10.1088/1126-6708/2004/11/040}},
\href{http://www.arXiv.org/abs/hep-ph/0409146}{\texttt{arXiv:hep-ph/0409146}}.

\bibitem{HERWIG}
\hrefCMSnoop {}{G.~Corcella {et~al.}, ``{HERWIG 6}: An event generator for
  hadron emission reactions with interfering gluons (including supersymmetric
  processes)'',} \textit{ JHEP} \textbf{ 01} (2001) 010,
  \href{http://dx.doi.org/10.1088/1126-6708/2001/01/010}{\doi{10.1088/1126-6708/2001/01/010}},
  \href{http://www.arXiv.org/abs/hep-ph/0011363}{\texttt{arXiv:hep-ph/0011363}}.

\bibitem{auet2tune}
\href {http://cdsweb.cern.ch/record/1363300}{{ATLAS Collaboration}, ``{ATLAS
  tunes of PYTHIA 6 and Pythia 8 for MC11}'',} ATLAS PUB note
  ATL-PHYS-PUB-2011-009, 2011.

\bibitem{CT10}
\hrefCMSnoop {}{L.~Hung-Liang {et~al.}, ``{New parton distributions for
  collider physics}'',} \textit{ Phys. Rev. D} \textbf{ 82} (2010) 074024,
  \href{http://dx.doi.org/10.1103/PhysRevD.82.074024}{\doi{10.1103/PhysRevD.82.074024}},
  \href{http://www.arXiv.org/abs/hep-ph/1007.2241}{\texttt{arXiv:hep-ph/1007.2241}}.

\bibitem{Cacciari:2005hq}
\hrefCMSnoop {}{M.~Cacciari and G.~P. Salam, ``{Dispelling the $N^{3}$ myth for
  the $k_t$ jet-finder}'',} \textit{ Phys. Lett. B} \textbf{ 641} (2006) 57,
  \href{http://dx.doi.org/10.1016/j.physletb.2006.08.037}{\doi{10.1016/j.physletb.2006.08.037}},
\href{http://www.arXiv.org/abs/hep-ph/0512210}{\texttt{arXiv:hep-ph/0512210}}.

\bibitem{Cacciari:2008gp}
\hrefCMSnoop {}{M.~Cacciari, G.~P. Salam, and G.~Soyez, ``{The anti-$k_t$ jet
  clustering algorithm}'',} \textit{ JHEP} \textbf{ 04} (2008) 063,
  \href{http://dx.doi.org/10.1088/1126-6708/2008/04/063}{\doi{10.1088/1126-6708/2008/04/063}},
\href{http://www.arXiv.org/abs/0802.1189}{\texttt{arXiv:0802.1189}}.

\bibitem{CMS-PAS-PFT-09-001}
\href {http://cdsweb.cern.ch/record/1194487}{{CMS Collaboration},
  ``Particle--Flow Event Reconstruction in {CMS} and Performance for Jets,
  Taus, and {\MET}'',} CMS Physics Analysis Summary CMS-PAS-PFT-09-001, 2009.

\bibitem{CMS-PAS-PFT-10-001}
\href {http://cdsweb.cern.ch/record/1247373}{{CMS Collaboration},
  ``Commissioning of the Particle-flow Event Reconstruction with the first
  {LHC} collisions recorded in the {CMS} detector'',} CMS Physics Analysis
  Summary CMS-PAS-PFT-10-001, 2010.

\bibitem{JES}
\hrefCMSnoop {}{{CMS Collaboration}, ``Determination of jet energy calibration
  and transverse momentum resolution in {CMS}'',} \textit{ J. Instrum.}
  \textbf{ 6} (2011) P11002,
  \href{http://dx.doi.org/10.1088/1748-0221/6/11/P11002}{\doi{10.1088/1748-0221/6/11/P11002}}.

\bibitem{Chatrchyan:2012jua}
\hrefCMSnoop {}{{CMS Collaboration}, ``{Identification of b-quark jets with the
  CMS experiment}'',} \textit{ JINST} \textbf{ 8} (2013) P04013,
  \href{http://dx.doi.org/10.1088/1748-0221/8/04/P04013}{\doi{10.1088/1748-0221/8/04/P04013}},
\href{http://www.arXiv.org/abs/1211.4462}{\texttt{arXiv:1211.4462}}.

\bibitem{Agashe:2014kda}
\hrefCMSnoop {}{{Particle Data Group} Collaboration, ``{Review of Particle
  Physics}'',} \textit{ Chin. Phys. C} \textbf{ 38} (2014) 090001,
\href{http://dx.doi.org/10.1088/1674-1137/38/9/090001}{\doi{10.1088/1674-1137/38/9/090001}}.

\bibitem{Chatrchyan:2013xza}
\hrefCMSnoop {}{{CMS Collaboration}, ``{Measurement of the top-quark mass in
  all-jets $\ttbar$ events in pp collisions at $\sqrt{s}=7$ TeV}'',} \textit{
  Eur. Phys. J. C} \textbf{ 74} (2014) 2758,
  \href{http://dx.doi.org/10.1140/epjc/s10052-014-2758-x}{\doi{10.1140/epjc/s10052-014-2758-x}},
\href{http://www.arXiv.org/abs/1307.4617}{\texttt{arXiv:1307.4617}}.

\bibitem{CMS-PAS-LUM-13-001}
\href {http://cdsweb.cern.ch/record/1598864}{{CMS Collaboration}, ``CMS
  Luminosity Based on Pixel Cluster Counting - Summer 2013 Update'',} CMS
  Physics Analysis Summary CMS-PAS-LUM-13-001, 2013.

\bibitem{Alekhin:2011sk}
\hrefCMSnoop {}{S.~Alekhin {et~al.}, ``{The PDF4LHC Working Group Interim
  Report}'',} (2011).
\href{http://www.arXiv.org/abs/1101.0536}{\texttt{arXiv:1101.0536}}.

\bibitem{Botje:2011sn}
\hrefCMSnoop {}{M.~Botje {et~al.}, ``{The PDF4LHC Working Group Interim
  Recommendations}'',} (2011).
\href{http://www.arXiv.org/abs/1101.0538}{\texttt{arXiv:1101.0538}}.

\bibitem{Skands:2010ak}
\hrefCMSnoop {}{P.~Z. Skands, ``{Tuning Monte Carlo Generators: The Perugia
  Tunes}'',} \textit{ Phys. Rev. D} \textbf{ 82} (2010) 074018,
  \href{http://dx.doi.org/10.1103/PhysRevD.82.074018}{\doi{10.1103/PhysRevD.82.074018}},
\href{http://www.arXiv.org/abs/1005.3457}{\texttt{arXiv:1005.3457}}.

\bibitem{Bayes}
\hrefCMSnoop {}{G.~D'Agostini, ``A multidimensional unfolding method based on
  Bayes' theorem'',} \textit{ Nucl. Instrum. Meth. A} \textbf{ 362} (1995) 487,
  \href{http://dx.doi.org/10.1016/0168-9002(95)00274-X}{\doi{10.1016/0168-9002(95)00274-X}}.

\bibitem{Adye:2011gm}
\href {http://inspirehep.net/record/898599/files/arXiv:1105.1160.pdf}{T.~Adye,
  ``{Unfolding algorithms and tests using RooUnfold}'',} in \textit{
  {Proceedings of the PHYSTAT 2011 Workshop, CERN, Geneva, Switzerland, January
  2011, CERN-2011-006}}, p.~313.
\newblock 2011.
\newblock
\href{http://www.arXiv.org/abs/1105.1160}{\texttt{arXiv:1105.1160}}.
\newblock

\bibitem{Czakon:2011xx}
\hrefCMSnoop {}{M.~Czakon and A.~Mitov, ``{Top++}: A program for the
  calculation of the top-pair cross-section at hadron colliders'',} \textit{
  Comput. Phys. Commun.} \textbf{ 185} (2014) 2930,
  \href{http://dx.doi.org/10.1016/j.cpc.2014.06.021}{\doi{10.1016/j.cpc.2014.06.021}},
\href{http://www.arXiv.org/abs/1112.5675}{\texttt{arXiv:1112.5675}}.

\bibitem{Czakon:2013goa}
\hrefCMSnoop {}{M.~Czakon, P.~Fiedler, and A.~Mitov, ``Total Top-Quark
  Pair-Production Cross Section at Hadron Colliders Through
  {$\mathcal{O}(\alpha^{4}_{S})$}'',} \textit{ Phys. Rev. Lett.} \textbf{ 110}
  (2013) 252004,
  \href{http://dx.doi.org/10.1103/PhysRevLett.110.252004}{\doi{10.1103/PhysRevLett.110.252004}},
\href{http://www.arXiv.org/abs/1303.6254}{\texttt{arXiv:1303.6254}}.

\end{thebibliography}\endgroup
